\documentclass[12pt,a4paper,final]{iopart}

\usepackage{iopams}

\usepackage{graphicx}% Include figure files
\usepackage{dcolumn}% Align table columns on decimal point
\usepackage[usenames,dvipsnames,svgnames,table]{xcolor}
\usepackage[colorlinks=true,
            linkcolor=blue,
            urlcolor=black,
            citecolor=blue]{hyperref}
\usepackage{bm}% bold math
\usepackage{enumerate}
\usepackage{lipsum}
\usepackage{epstopdf}
\usepackage{float}
\usepackage[caption = false]{subfig}
\usepackage{tabularx}
\usepackage{color}
\begin{document}

\title{Density profiles of two-component Bose-Einstein condensates  interacting with a Laguerre-Gaussian Beam}% Force line breaks with \\

\author[cor1]{Anal Bhowmik$^{1}$, Pradip Kumar Mondal$^2$, Sonjoy Majumder$^1$, and  Bimalendu Deb$^3$}
\address{$^1$Department of Physics, Indian Institute of Technology Kharagpur, Kharagpur-721302, India.}

\address{$^2$Department of Physics, Egra Sarada Shashi Bhusan  College, Egra-721429, India.}

\address{$^3$Department of Materials Science, Indian Association for the Cultivation of Science, Jadavpur, Kolkata 700032, India.}

\eads{\mailto{analbhowmik@phy.iitkgp.ernet.in}}

\begin{abstract}
The density profiles of trapped two-component Bose-Einstein condensates (BEC) and its microscopic interaction with Laguerre Gaussian (LG) beam are studied.  We consider the $^{87}$Rb BEC in two hyperfine spin  components. The wavelength of the LG beam is assumed to be comparable to the atomic de-Broglie wavelength. Competitions between intra- and inter-component interactions produce interesting density structures of the ground state of  BEC.    We demonstrate  vortex-antivortex interference and its dependence on the inter-component interactions and Raman transitions.

\end{abstract}


\section{INTRODUCTION}

Binary mixtures of Bose-Einstein condensates (BEC) of  different atomic species have been the subject of  intensive theoretical  and experimental research \cite{Hall1998, Myatt1997, Kurn1998, Maddaloni2000, Matthews1999, Anderson2000, Ho1996, Jezek2001, Jezek2005, Chui2000, Chui2001, Chui2002,  Law1997, Ao1998, Timmermans1998,  Martin1999,  Trippenbach2000}. The mixtures can be composed of  two different alkali-metal atomic gases \cite{Modugno2002, Thalhammer2008, McCarron2011,  Wacker2015, Wang2016, Lercher2011, Pasquiou2013, Roy2015, Lee2016} or two different isotopes of same element \cite{Papp2008, Sugawa2011, Inouye1998,  Tojo2010} or   same element with different hyperfine states  \cite{Hall1998, Myatt1997, Kurn1998,  Maddaloni2000,  Stenger1998,  Sadler2006} etc.
These mixtures provide a unique opportunity for exploring  fascinating many-body quantum physics that can not be studied with a single-component Bose-Einstein condensate.  For instance, phase separation \cite{ McCarron2011, Wacker2015, Wang2016, Papp2008}, pattern formation \cite{Sabbatini2011, Hoefer2011, Hamner2011, De2014}, symmetry breaking
transitions \cite{Lee2009}, skyrmions \cite{Kawakami2012}, collective modes \cite{ Maddaloni2000, Barbut2014}, nonlinear dynamical
excitations \cite{Mertes2007, Eto2016}, quantum turbulence \cite{Takeuchi2010} and vortex bright solitons \cite{Law2010} have been studied with the binary BEC.  Further, a separated or resolved phase of a two-component BEC is  essential  for   Kelvin-Helmholtz  \cite{Suzuki2010}
and  Rayleigh-Taylor instability \cite{Sasaki2009}.

  Over the years,  there have been many studies of the density structure of two-component BEC, but in all the  cases, analyses were limited close to the center  of the trap \cite{ McCarron2011, Wacker2015, Wang2016}.  In contrast, the density structures of two-component BEC are important when the BEC interacts with an LG beam far away from the trap center.  In other words, the interaction with the LG beam may provide a means  to  study the matter density distributions away from the trap axis.  This interaction generates quantized vortices in BEC either through Raman processes \cite{Marzlin1997, Nandi2004, Mondal2015, Bhowmik2016} or slow light consideration \cite{Dutton2004}. Because of the meandered structure of the components, our calculations show the  criss-cross behavior of two-photon Rabi frequencies for Raman transitions with increasing inter-species coupling.  This behaviour provides contrast in phase density in the final state after the Raman transitions, when two-counter propagating LG beams  with proper frequencies interact simultaneously at individual components. This can be observed in the interference  of the vortices generated in the interaction with varying  coupling strengths.   Moreover, for larger values of the orbital angular momentum (OAM)  of light, the interaction will largely depend on the peripheral density profile of the components.

 The  density phase contrast can be realized by generating a vortex-antivortex superposition of BEC. The   combination of vortex-antivortex matter-wave states, generated by the superposition of vortices of opposite circulation, exhibits interesting  petal-like interference structures \cite{ Bhowmik2016, Liu2006}  and intriguing dynamics \cite{Simula2008}.   The properties of  vortex-antivortex structure in binary BEC yield rich physics \cite{Wen2013, Kapale2005, Thanvanthri2008, Wen2010}. Kapale
\textit{et al.} \cite{Kapale2005} proposed a scheme to create a coherent superposition of vortex-antivortex  in multi-component BECs using the optical vortex. They varied the population of vortex and antivortex states by changing the two-photon detuning parameter. 
The   superposition   resembles  the counter-rotating persistent currents in superconducting circuits \cite{Nakamura1999, Friedman2000, Wal2000} which are promising candidates for qubits in quantum-information processing and quantum communication networks \cite{Spedalieri2006}.

In this paper, we develop a theory for the interaction of Laguerre Gaussian (LG) beam with binary mixtures of the BEC and investigate the   variation in the superposition of vortex-antivortex structure using  two-photon Raman technique.   To obtain the interference pattern  of vortex-antivortex states, we have calculated the Rabi frequency of two-photon stimulated Raman transitions.  Interesting physics can be   investigated here by analyzing the  mixtures of components of BECs and their inter-component interaction strength.  For example,   the population of vortex  states in BEC components can be tuned by changing   the inter-component interaction strength. 

This paper is organized as follows. In Sec. II, we discuss the theory behind the generation of vortex state in binary mixtures of BEC.  In Sec. III, we study the density profiles of the ground state in detail. Also, the variation of Rabi frequency with  inter-component  interaction, number of particles, and the intensity of trapping potential are presented in the same section. In Sec IV, we show the change in the superposition of  vortex-antivortex state for the  two-component BEC of equal and unequal number of topological charges.  The conclusion is outlined in Sec. V.

\section{THEORY}

 A dilute mixture of two components of a BEC trapped in a harmonic potential is considered here. To describe the stationary ground-state of this system at  zero temperature limit, one can use the coupled Gross-Pitaevskii (GP) equations \cite{Ho1996, Jezek2001, Pu1997, Kevrekidis2008} in cylindrical coordinate (see APPENDIX), 

\begin{equation}
\hspace{-2.3cm}\left[-\frac{\hbar^2\nabla^2}{2m_1}+\frac{1}{2}m_1(\omega^2_{\bot}R^2+\omega^2_{Z}Z^2)+\frac{\kappa^2}{R^2} + U_{11}|\Psi_1(\textbf{R})|^2+ U_{12}|\Psi_2(\textbf{R})|^2\right]\Psi_1(\textbf{R})=\mu_1 \Psi_1(\textbf{R}),
\end{equation}

 \begin{equation}
\hspace{-2.3cm}\left[-\frac{\hbar^2\nabla^2}{2m_2}+\frac{1}{2}m_2(\omega^2_{\bot}R^2+\omega^2_{Z}Z^2)+\frac{\kappa^2}{R^2}+ U_{22}|\Psi_2(\textbf{R})|^2+U_{21}|\Psi_1(\textbf{R})|^2\right]\Psi_2(\textbf{R})=\mu_2 \Psi_2(\textbf{R})
\end{equation}
with the  normalization condition $\int |\Psi_i(\textbf{R})|^2 d\textbf{R}=N_i$. Here $N_i$, $m_i$ and $\mu_i$ denote the number of atoms, mass of the atom, and the chemical potential of the $i$-th ($i$=1 \& 2) component of BEC. $\kappa$ is the quantum of circulation of atoms about the $z$ axis. $\Psi_1$ and $\Psi_2$ are the center-of-mass (CM) wavefunctions of the components, say, BEC-1 and BEC-2.  $\omega_{\bot}$ and $\omega_Z$ are trapping potentials in the  $x-y$ plane and along the $z$ axis, respectively. $U_{11}$ and $U_{22}$ are the intra-component coupling strengths of species 1  and 2, respectively. $U_{12}$ and $U_{21}$ are the inter-component coupling strengths between the species. These coupling strengths are related to intra- and inter-component $s$-wave scattering lengths via the relations, $U_{11}={4\pi a_{11} \hbar^2}/{m_1}$, $U_{12}=U_{21}={2\pi a_{12} \hbar^2}(m_1+m_2)/{m_1m_2}$ and $U_{22}={4\pi a_{22} \hbar^2}/{m_2}$. Now atoms in each of the BEC components are considered to be of  the simplest form, a valance electron of  charge $-e$ and mass $m_e$ roaming  around core electron and nucleus of total charge $+e$ and mass $m_n$. The CM coordinate with respect to laboratory coordinate system  is $\textbf{R}=(m_e \textbf{r}_e + m_n \textbf{r}_n)/m_t $, where $m_t=m_e+m_n$ being the total mass.    Here $\textbf{r}_e$ and $\textbf{r}_n$ are the coordinates of the valance electron and the center of atom,  respectively, with respect to laboratory coordinate system and the relative (internal) coordinate can be expressed as $\textbf{r}=\textbf{r}_e -\textbf{r}_n$.

We consider the LG beam without any off-axis node, propagating along the $z$ axis of the laboratory frame. The beam  interacts with the coupled BEC whose de Broglie wavelength is large enough to feel the intensity variation of the LG beam but smaller than the waist of the beam. Let $ \psi_i$ and $\Psi_i $ be the internal (electronic)  and the CM wavefunction, respectively, of $i $-th component of BEC. Then the total wavefunction of the system of two-component BEC can be written as, $\Upsilon(\textbf{R}_1,  \textbf{R}_2, \textbf{r}_1, \textbf{r}_2)=\Psi_1({\textbf{R}_1}) \Psi_2({\textbf{R}_2})\psi_1({\textbf{r}_1}) \psi_2({\textbf{r}_2})$. Here the atom-radiation interaction Hamiltonian, $H_{int}$,  is derived from the Power-Zienau-Wooley (PZW) scheme \cite{Babiker2002}, which is beyond the level of dipole approximation.

\begin{equation}
H_{int}=-\int d\textbf{r}^\prime P(\textbf{r}^\prime)\boldsymbol{.} \textbf{E}(\textbf{r}^\prime, t) +h.c.
\end{equation}
where $\textbf{E}(\textbf{r}^\prime, t)$ is the local electric field of the LG beam \cite{Mondal2014, Mukherjee2018} experienced by the atom. $P(\textbf{r}^\prime)$ is the electric polarization
given by
\begin{equation}
P(\textbf{r}^\prime)=-e\frac{m_n}{m_t}\textbf{r}\int_0^1 d\lambda \delta \Big(\textbf{r}^\prime-\textbf{R}-\lambda\frac{m_n}{m_t}\textbf{r}\Big).
\end{equation}

 If the LG beam interacts with one of the components of the BEC (say, $n$-th), then the dipole transition matrix element under paraxial approximation will be 

\begin{eqnarray}
\hspace{-2.3cm} M_{i \rightarrow f}^n  =\langle \Upsilon_f^i | H_{int} | \Upsilon _i \rangle  \nonumber
 &=&\sqrt{\frac{4\pi}{3 |l| !}}e\frac{m_n}{m_t}\sum_{\sigma=0,\pm 1}\epsilon_\sigma  \nonumber \\
 &\times & \Bigl [\langle \Psi _{nf}({\textbf{R}_n}) | \frac{R_n^{(|l|)}}{w_0^{|l|}}e^{il\Phi_n} e^{ikZ_n}| \Psi _{ni}({\textbf{R}_n}) \rangle \langle \psi _{nf}({\textbf{r}_n}) | r Y_1^{\sigma}(\boldsymbol{\hat{\textbf{r}}})| \psi _{ni}({\textbf{r}_n}) \rangle \nonumber \\
 &\times & \prod_{p\neq n}\langle \Psi _{pf}({\textbf{R}_p}) | \Psi _{pi}({\textbf{R}_p}) \rangle \langle \psi _{pf}({\textbf{r}_p}) | \psi _{pi}({\textbf{r}_p}) \rangle \Bigr],
\end{eqnarray}
where $\epsilon_\pm= (E_x \pm iE_y)/\sqrt{2}$ and  $\epsilon_0=E_z$. Eq. (5)  clearly shows that  the azimuthal coordinate ($\Phi_n$) of the CM  is changed by nucleation of vortex dictated by the topological charge of the beam. The polarization of the field interacts with the electronic motion, resulting in an electronic transition between the two internal states of the atoms. This  portion of the
transition matrix element is calculated using relativistic  coupled-cluster
theory \cite{Bhowmik2017a, Bhowmik2017b, Das2018}. Since both the components of the BEC are coupled  by an inter-component coupling, the creation of vortex in one of the components directly affects the wavefunction of  another  component of the BEC. But, here  we  consider  changes that occur only on the CM wavefunction of the latter component for the sake of  simplicity and the minor effect on their electronic motion is neglected.    

In the next section, we study numerical results of two-photon stimulated Raman transition using co-propagating LG and  Gaussian beams and discuss the variation of the Rabi frequencies  under the variation of inter-component coupling strength.
\begin{figure}[h]

\centering
\includegraphics[trim={0.9cm 0.9cm 0.01cm 0.01cm},width=7cm]{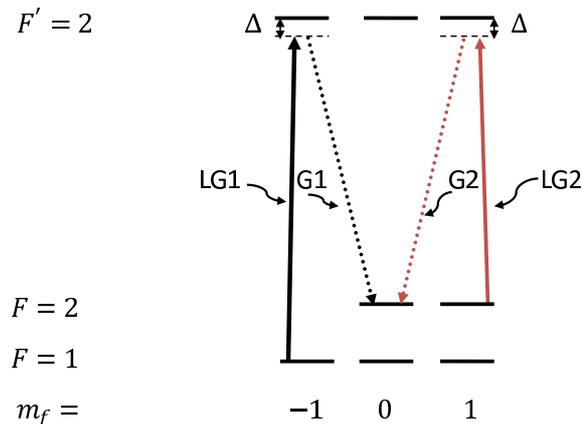}
\caption{Energy
level scheme of the two-photon Raman transitions. The atomic states show the  $^{87}$Rb hyperfine states. Atoms  are initially trapped in $| 5s_{\frac{1}{2}} F=1, m_f =-1 \rangle$ and $| 5s_{\frac{1}{2}} F=2, m_f =1 \rangle$. $\Delta$ represents two-photon detuning.}
\end{figure}

\section{NUMERICAL RESULTS AND INTERPRETATION}

 We consider that the  LG beam interacts with  a  coupled $ ^{87} $Rb BEC  prepared in $\psi_1=| 5S_{\frac{1}{2}}, F=1, m_f =-1 \rangle$ and $\psi_2=| 5S_{\frac{1}{2}}, F=2, m_f =1 \rangle$  hyperfine states  in a harmonic potential as discussed in the experimental work \cite{Hall1998}. For simplicity, we  consider that both the hyperfine states have been populated by   equal number of atoms.  We choose the characteristics of the experimental trap as given in Ref \cite{Andersen2006} with asymmetry parameter  $\lambda _{tr} =\omega _Z /\omega _\bot =2$ and the axial frequency $\omega _Z /2\pi =40$ Hz. The characteristic length is  $a _\bot =4.673$  $\mu$m. The intra-component $s$-wave scattering lengths are $a_{11}=1.03\times 5.5$nm, $a_{22}=0.97\times 5.5$nm \cite{Hall1998} and  the inter-component $s$-wave scattering length is $a_{12}=a_{21}=$ $g\times 5.5$ nm, where $g$ is a parameter which can be tuned  \cite{Inouye1998, Chin2010}. The intensity of the LG beam is $I =10^2 $ W cm$^{-2}$ and its waist $w _0 =10 ^{-4}$ m. 
 
 The interaction of the trapped atoms  with the LG beam has an extra physical degree of freedom than interaction with a Gaussian beam. The former interaction is expected to extract an extra physical feature of the BEC, like the orientation of vortex in the matter system \cite{Mondal2015}.  These features are dependent to a large extent on the interaction among the atoms.

Initially, we consider that  both the components of the BEC are in non-vortex states. Let the co-propagating LG and Gaussian (G) beams with appropriate polarization interact simultaneously with  BEC-1, using stimulated Raman transition to $|  5s_{\frac{1}{2}} F=2, m_f =0 \rangle$. The frequency difference between the two kinds of pulses, $\delta \nu_r$  is set equal to the recoil energy. The two-photon transitions are taken via $| 5p_{\frac{3}{2}}, F'=2, m_f =-1 \rangle$. Here G beam is detuned from the D2 line  by $\Delta=-1.5$ GHz ($\approx -150$ linewidths, enough to resist the destructive incoherent heating of the condensate due to spontaneous decay of excited states). Since the LG and G beams are co-propagating, the net transfer of linear momentum to the atom is zero. A similar case can be considered where the LG and G beams interact with  BEC-2 as shown in the Fig. 1. As the figure shows, both the cases of Raman transitions lead to the same final electronic state. Again, simultaneous application of the above two sets of LG and G beams to the corresponding components of BEC  lead to interference pattern at the final state, $| 5S_{\frac{1}{2}}, F=2, m_f =0 \rangle$. We would like to study the variation in  interference patterns of a vortex-antivortex pair with the inter-component interaction strength. To investigate  the interference patterns, we need to know the initial density profile of both  the components and their interaction with LG beam in terms of Rabi frequencies.

\begin{figure*}[!h]
\subfloat[]{\includegraphics[ trim = 7cm 1.5cm 7cm 2cm,scale=.20]{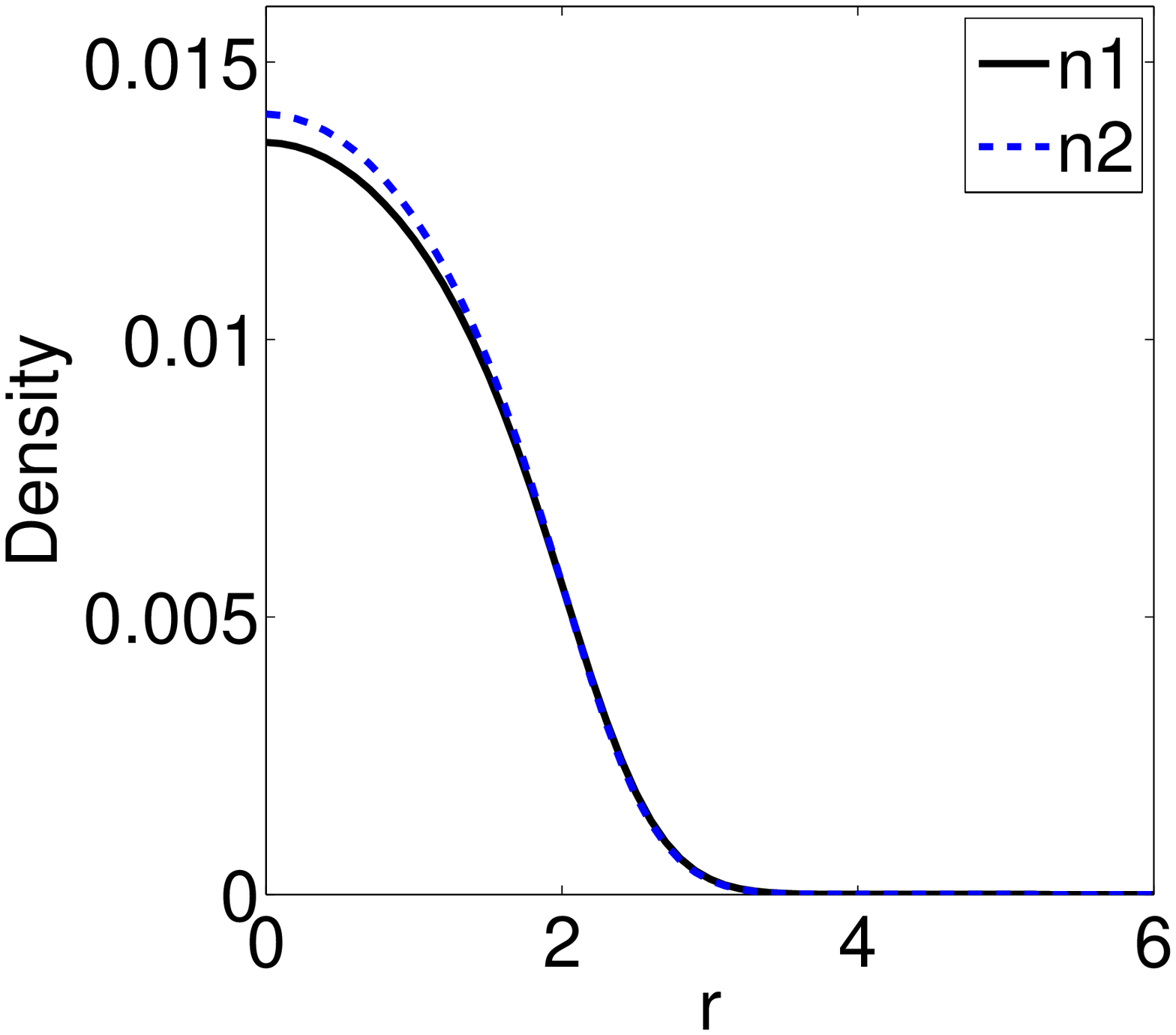}}\label{a}\hfil
\subfloat[]{\includegraphics[trim = 1cm 1.5cm 3cm 2cm,scale=.20]{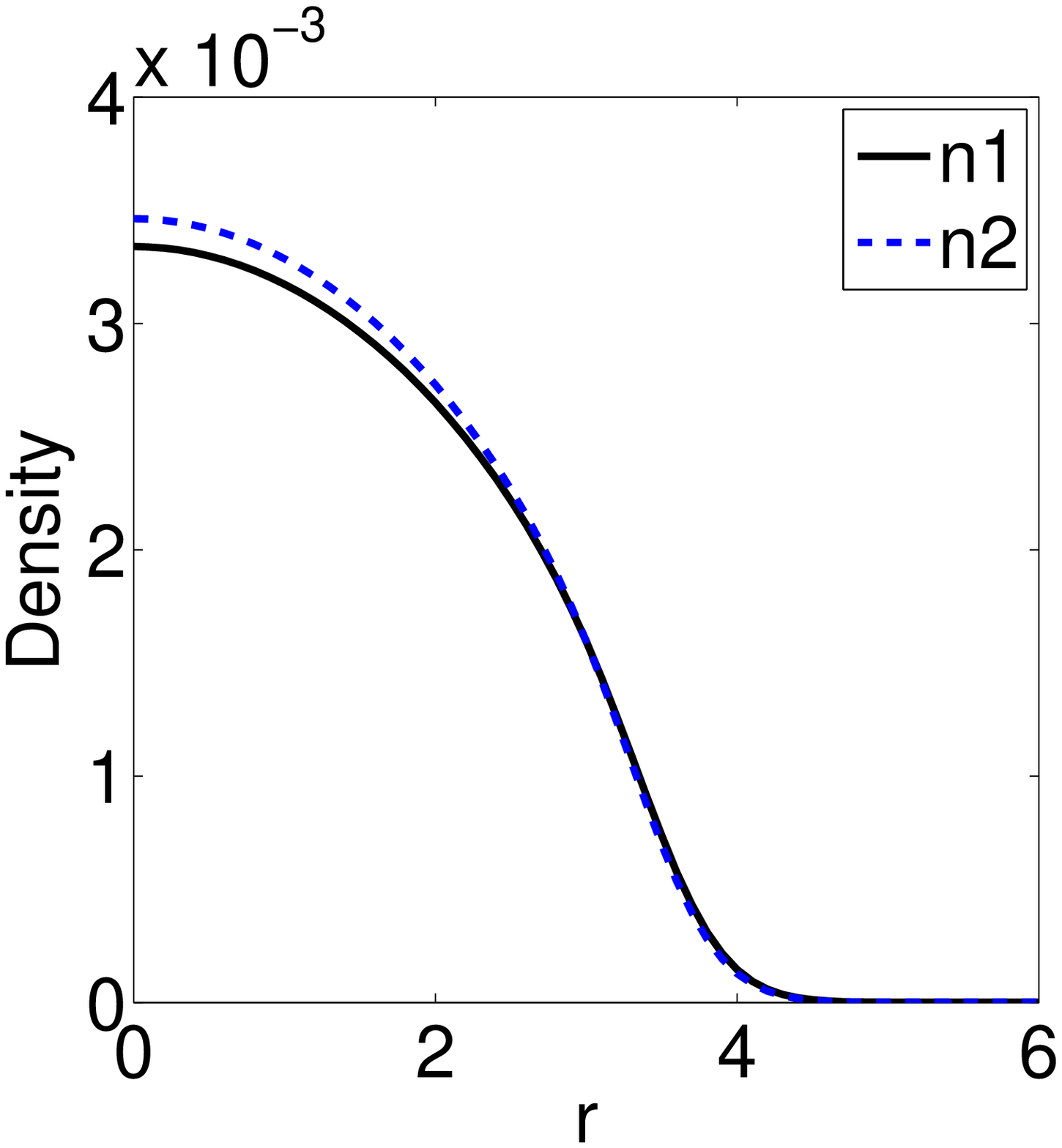}}\label{b}\hfil
\subfloat[]{\includegraphics[trim = 5cm 1.5cm 7cm 0cm,scale=.20]{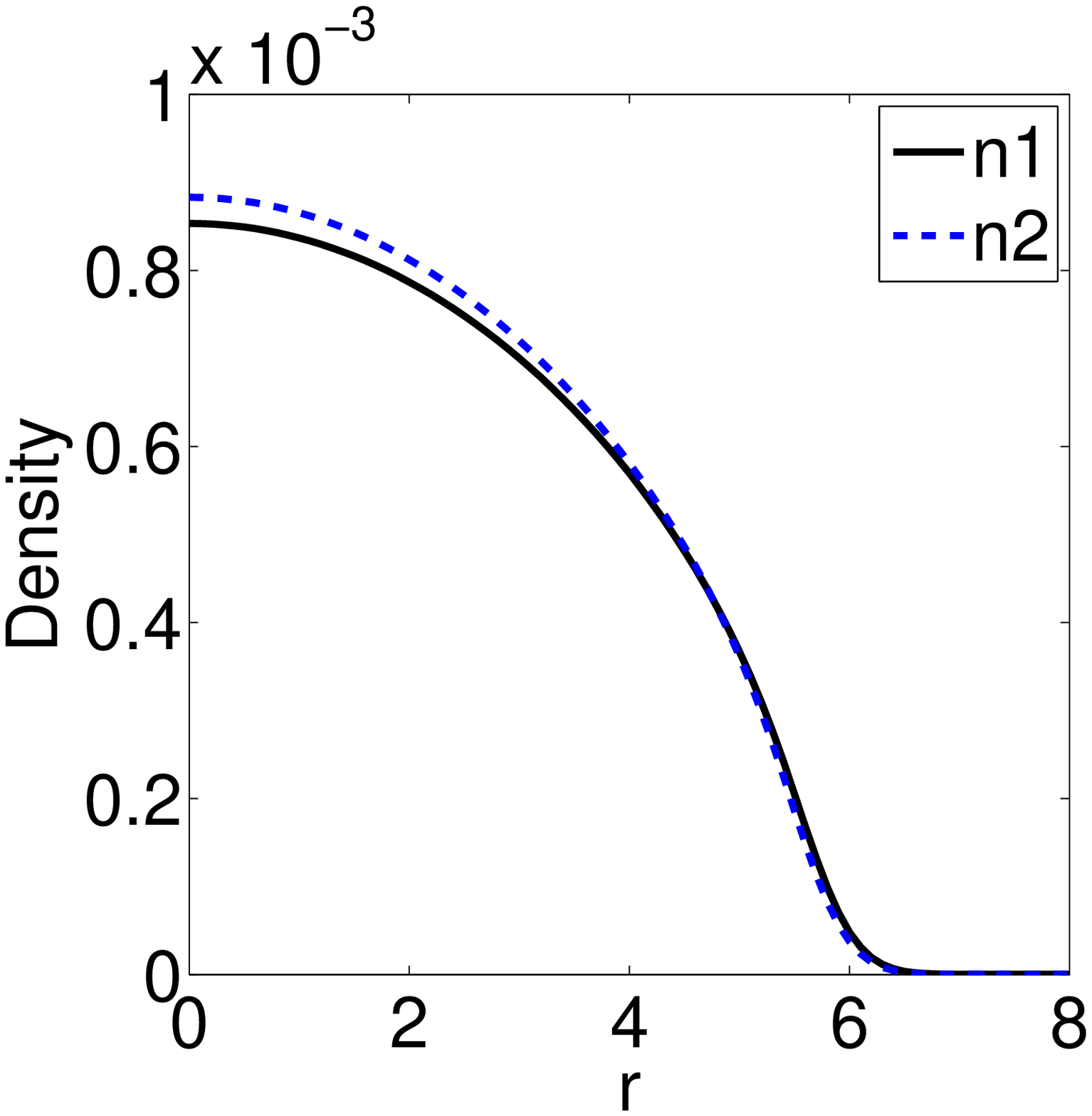}}\label{c}\\
\subfloat[]{\includegraphics[trim = 7cm 1.5cm 7cm 2cm,scale=.20]{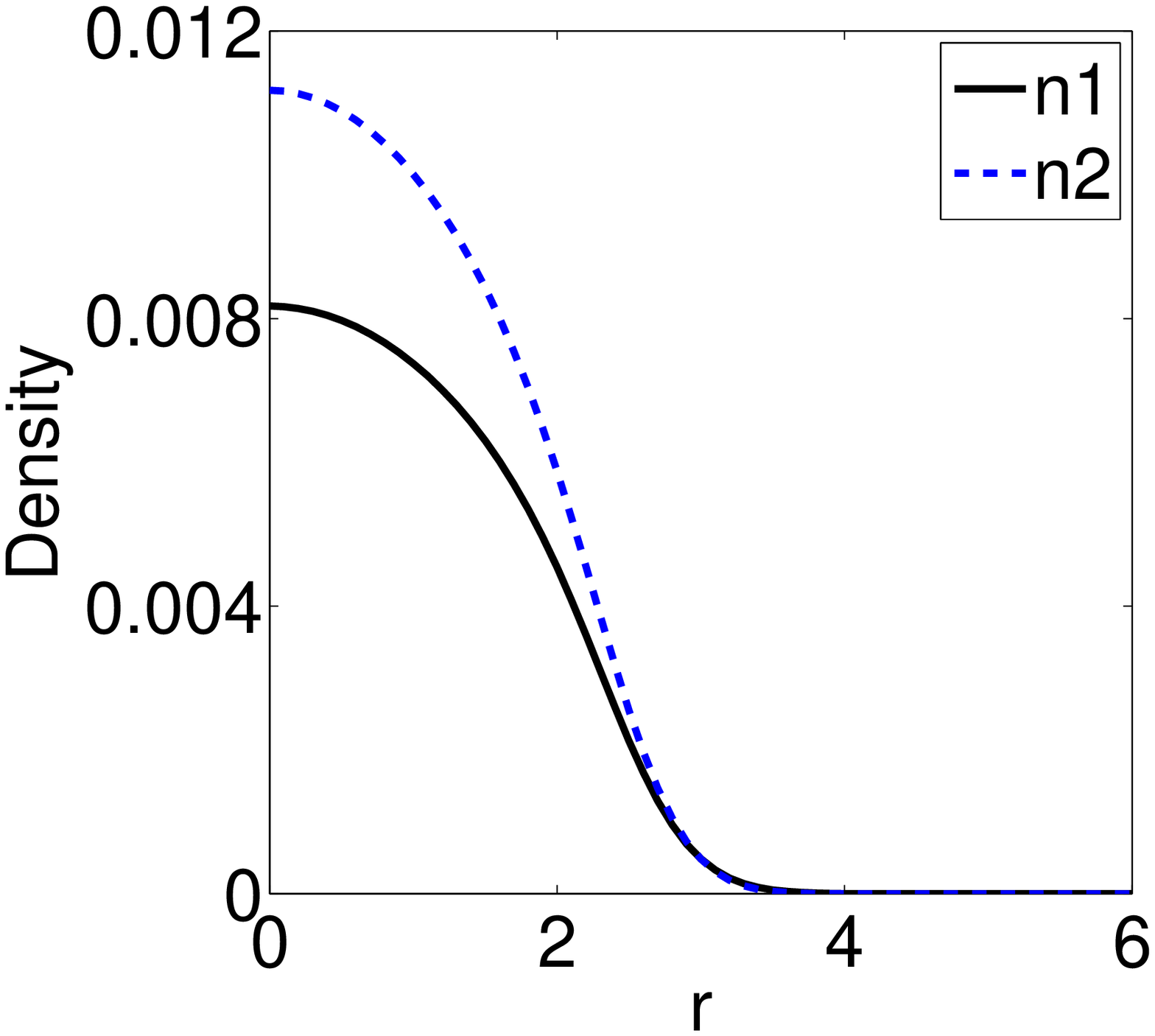}}\label{d}\hfil
\subfloat[]{\includegraphics[trim = 1cm 1.5cm 3cm 2cm,scale=.20]{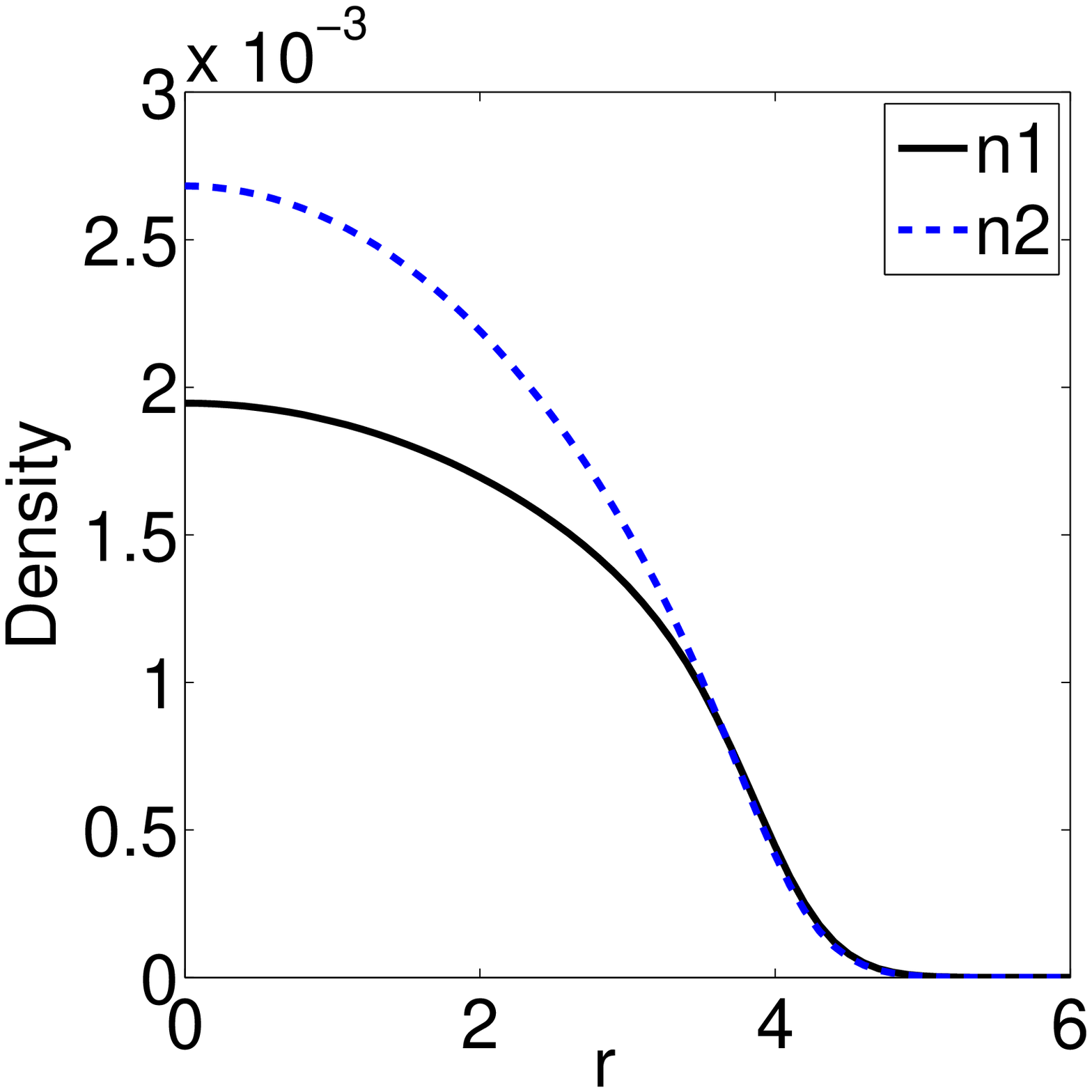}}\label{e}\hfil
\subfloat[]{\includegraphics[trim = 5cm 1.5cm 7cm 0cm,scale=.20]{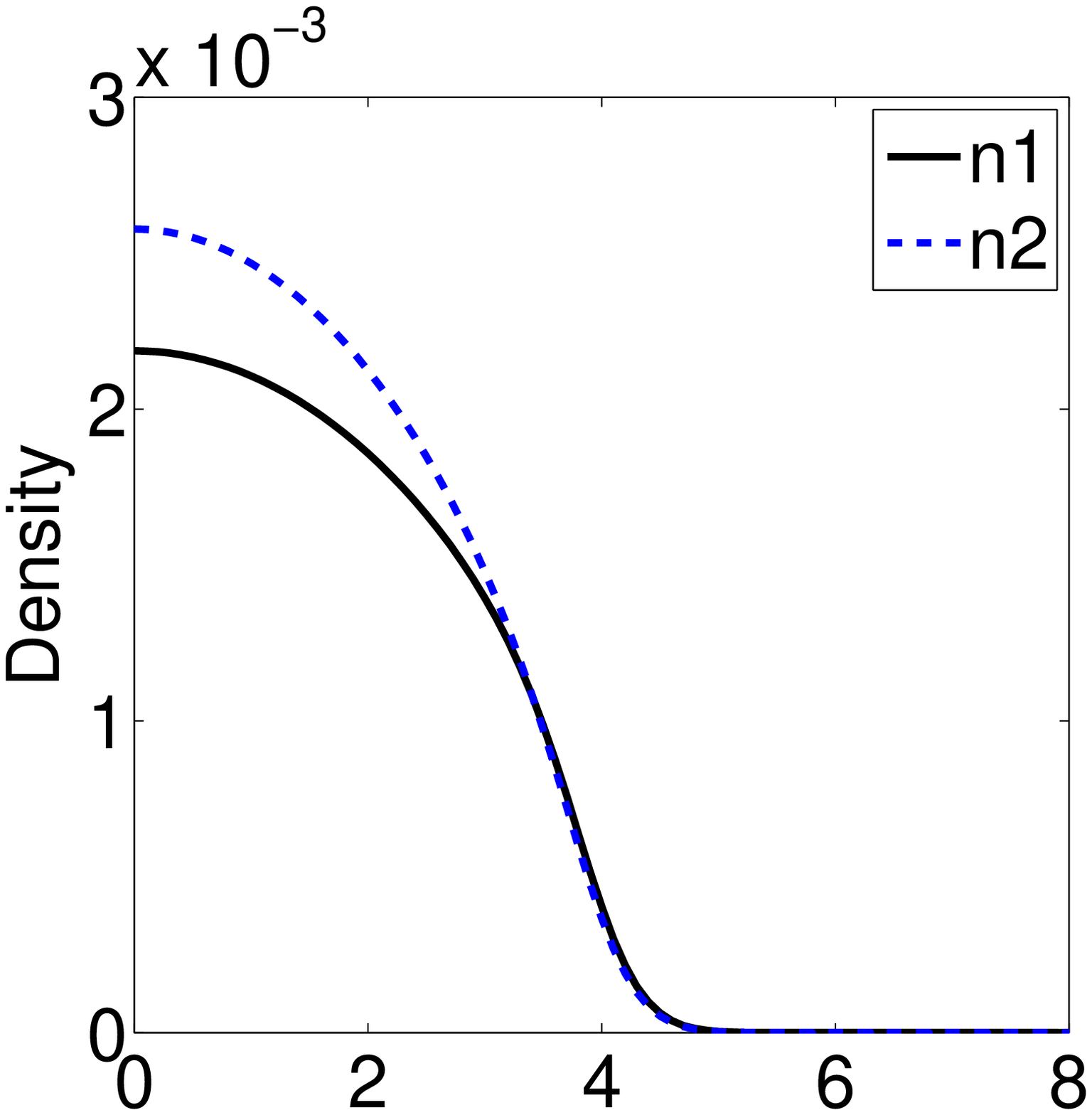}}\label{f}\\
\subfloat[]{\includegraphics[trim = 7cm 1.5cm 7cm 2cm,scale=.20]{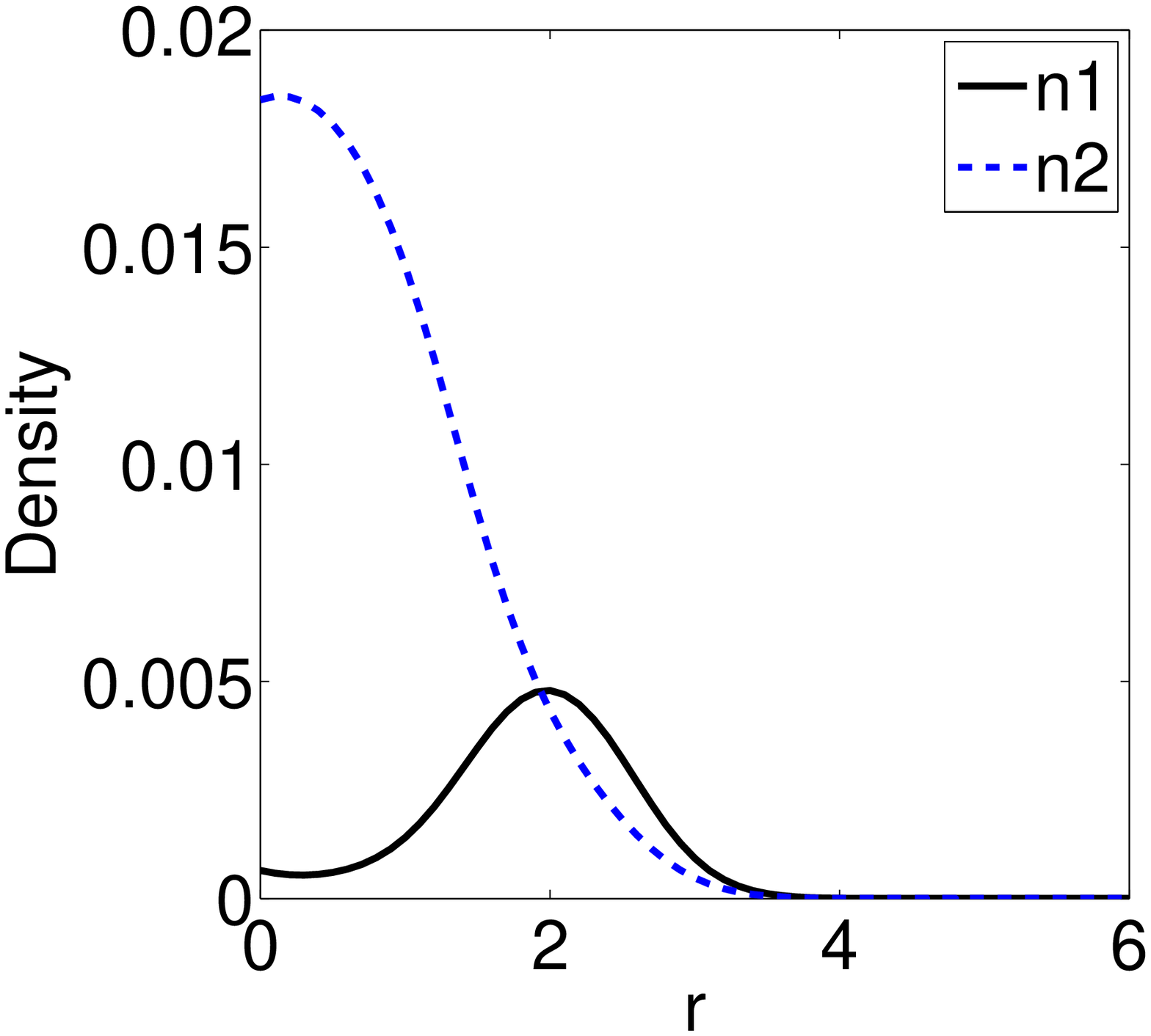}}\label{g}\hfil
\subfloat[]{\includegraphics[trim = 1cm 1.5cm 3cm 2cm,scale=.20]{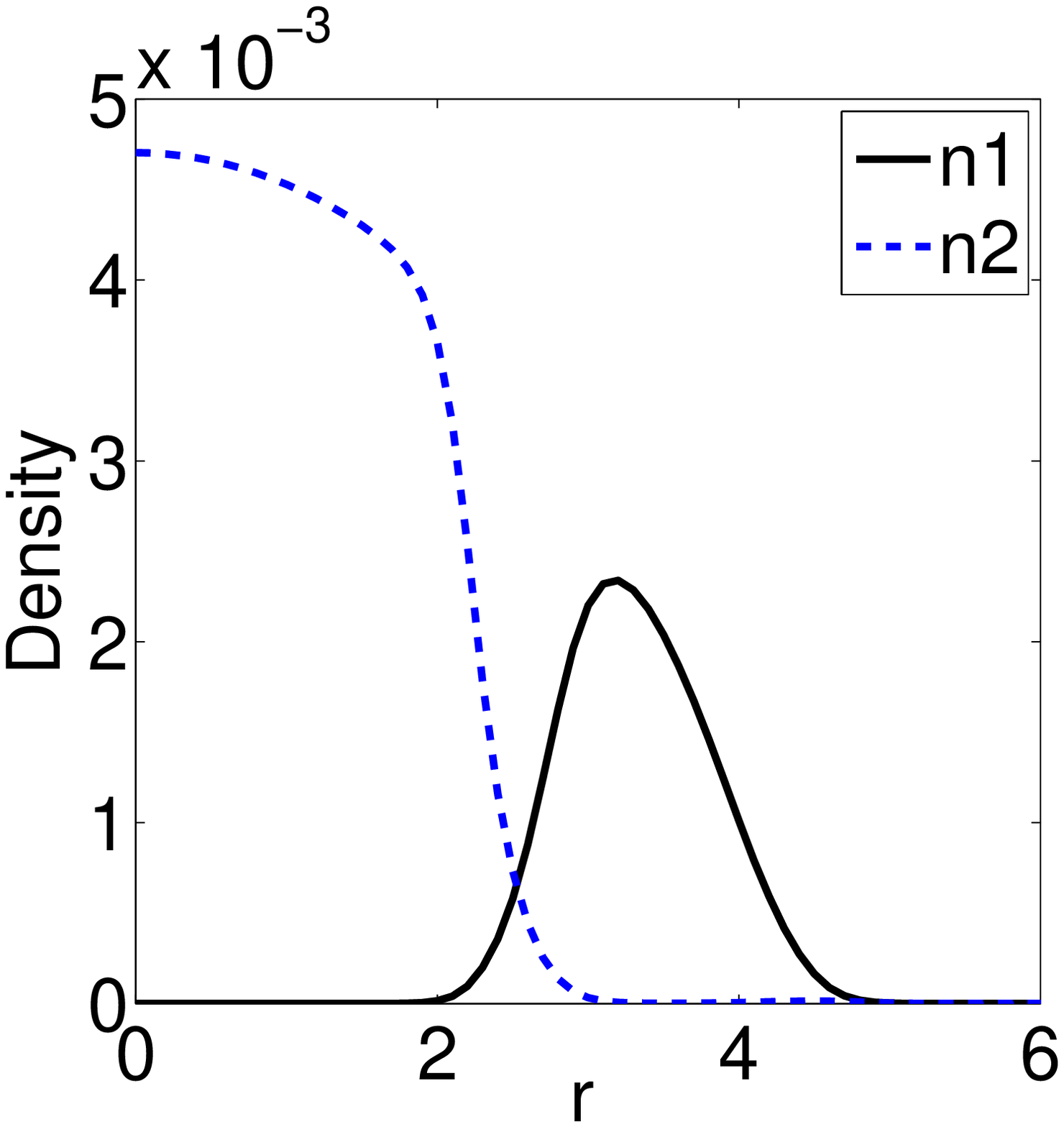}}\label{h}\hfil
\subfloat[]{\includegraphics[trim = 5cm 1.5cm 7cm 0cm,scale=.20]{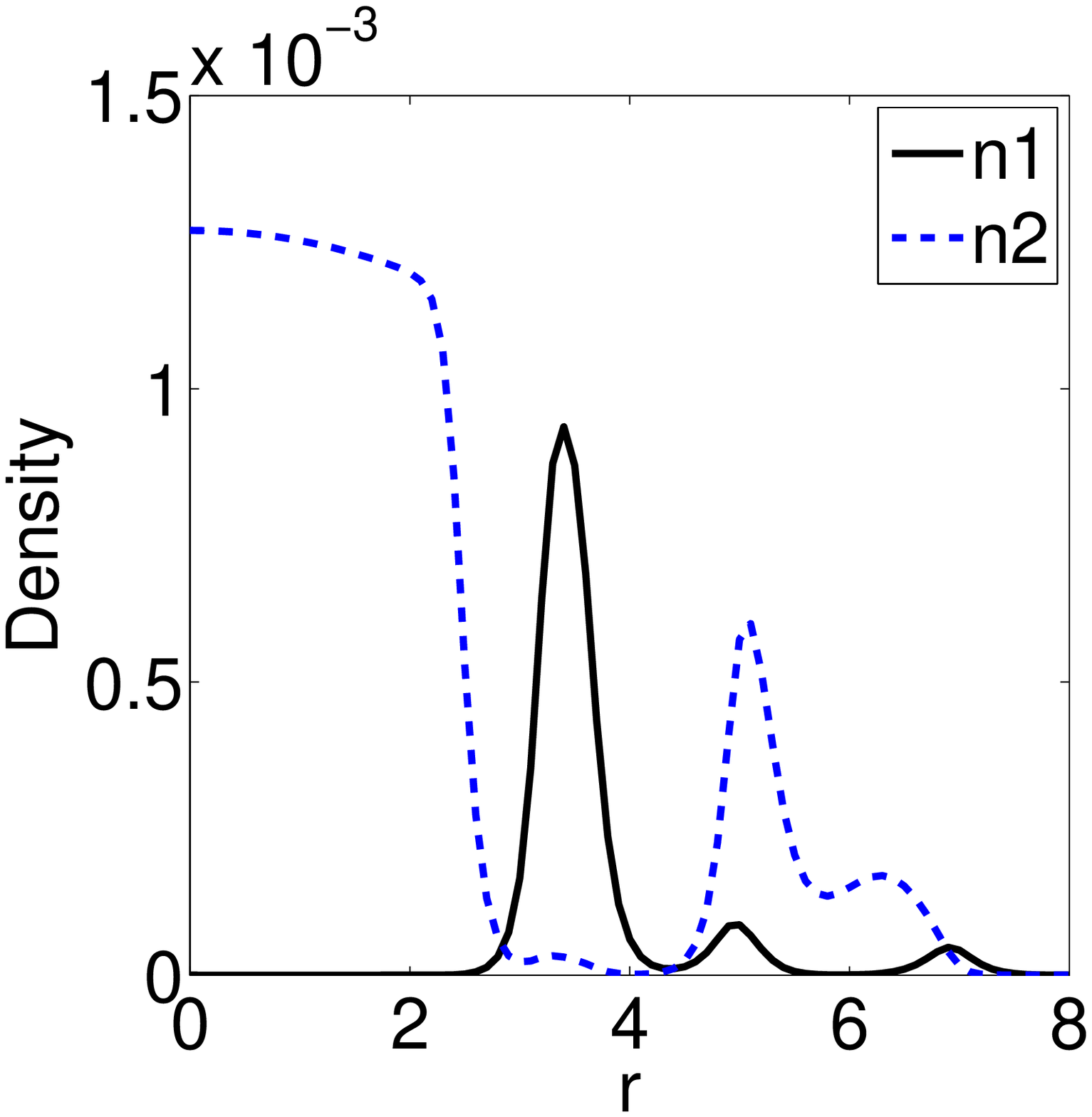}}\label{i}\\
\subfloat[]{\includegraphics[trim = 7cm 1.5cm 7cm 2cm,scale=.20]{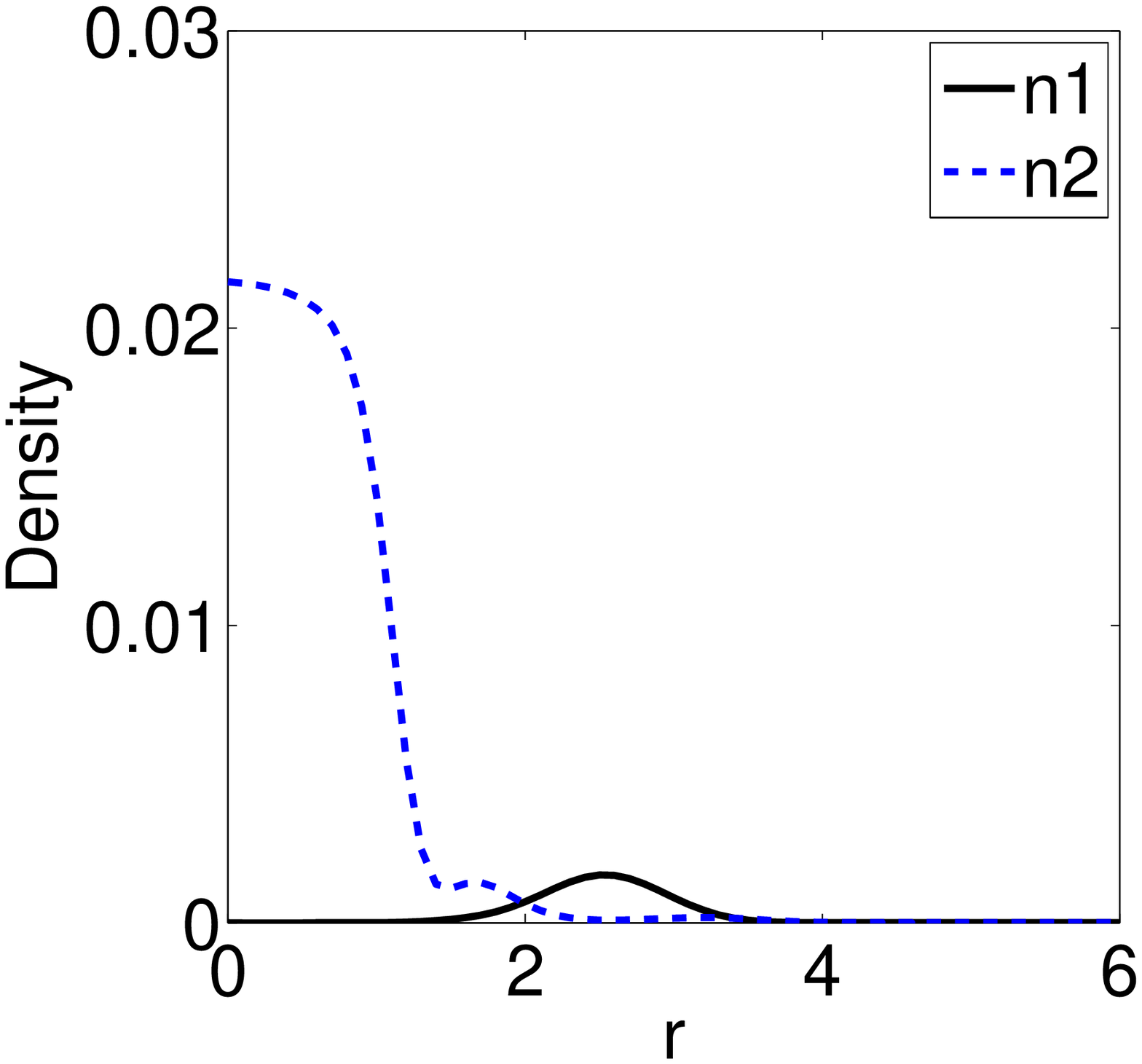}}\label{j}\hfil
\subfloat[]{\includegraphics[trim = 1cm 1.5cm 3cm 2cm,scale=.20]{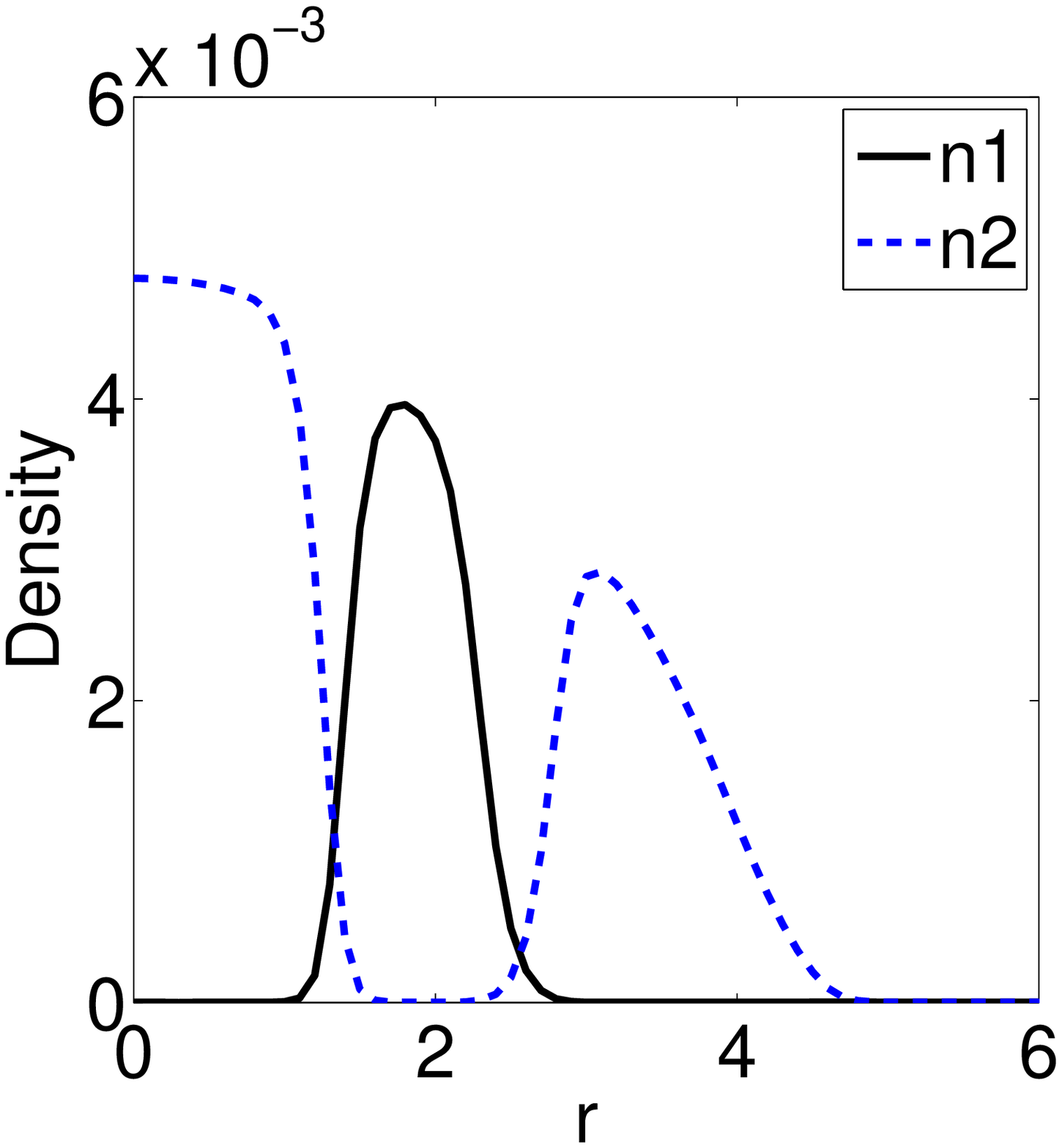}}\label{k}\hfil
\subfloat[]{\includegraphics[trim = 5cm 1.5cm 7cm 0cm,scale=.20]{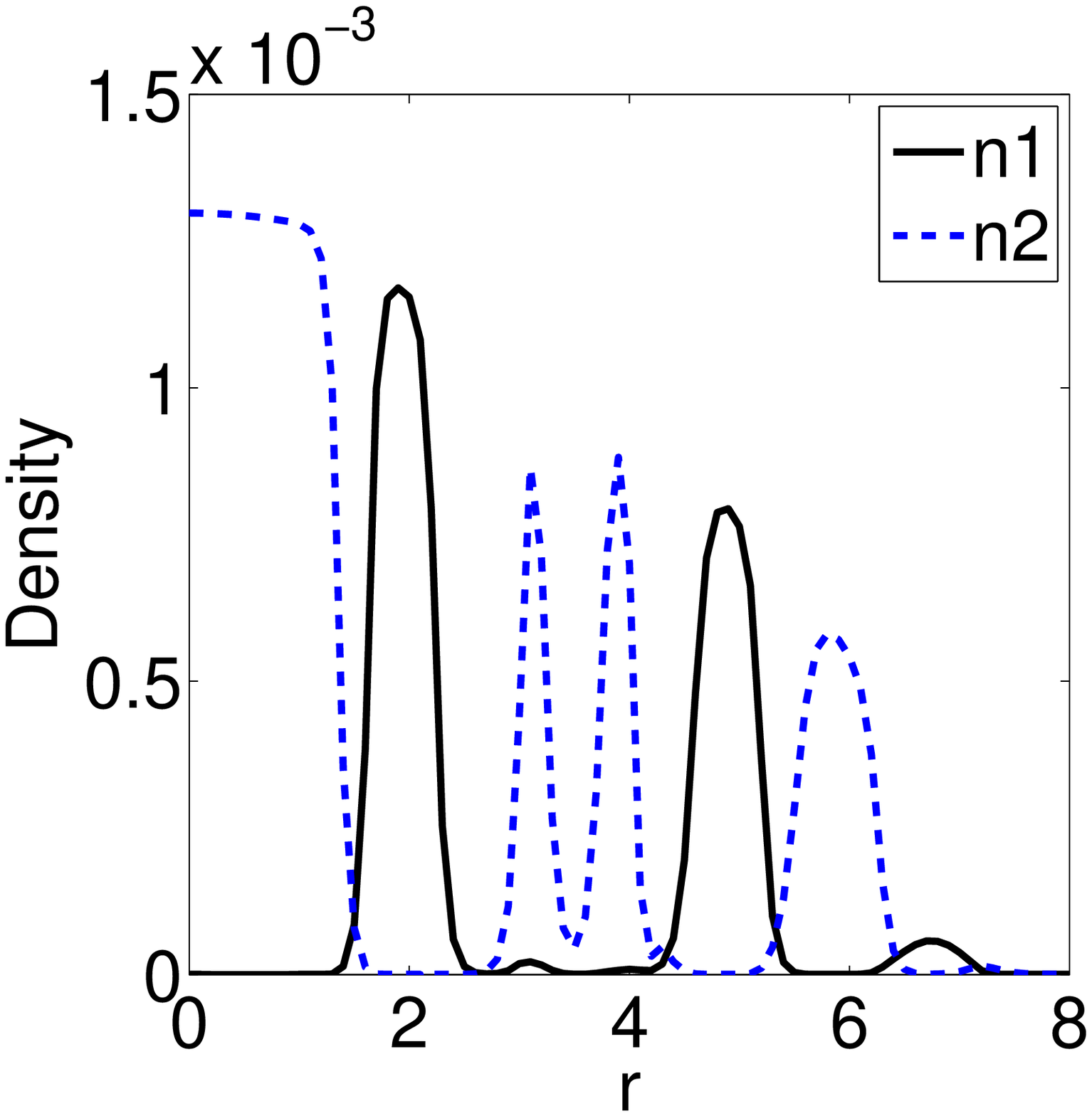}}\label{l}\\
\subfloat[]{\includegraphics[trim = 7cm 1.5cm 7cm 2cm,scale=.20]{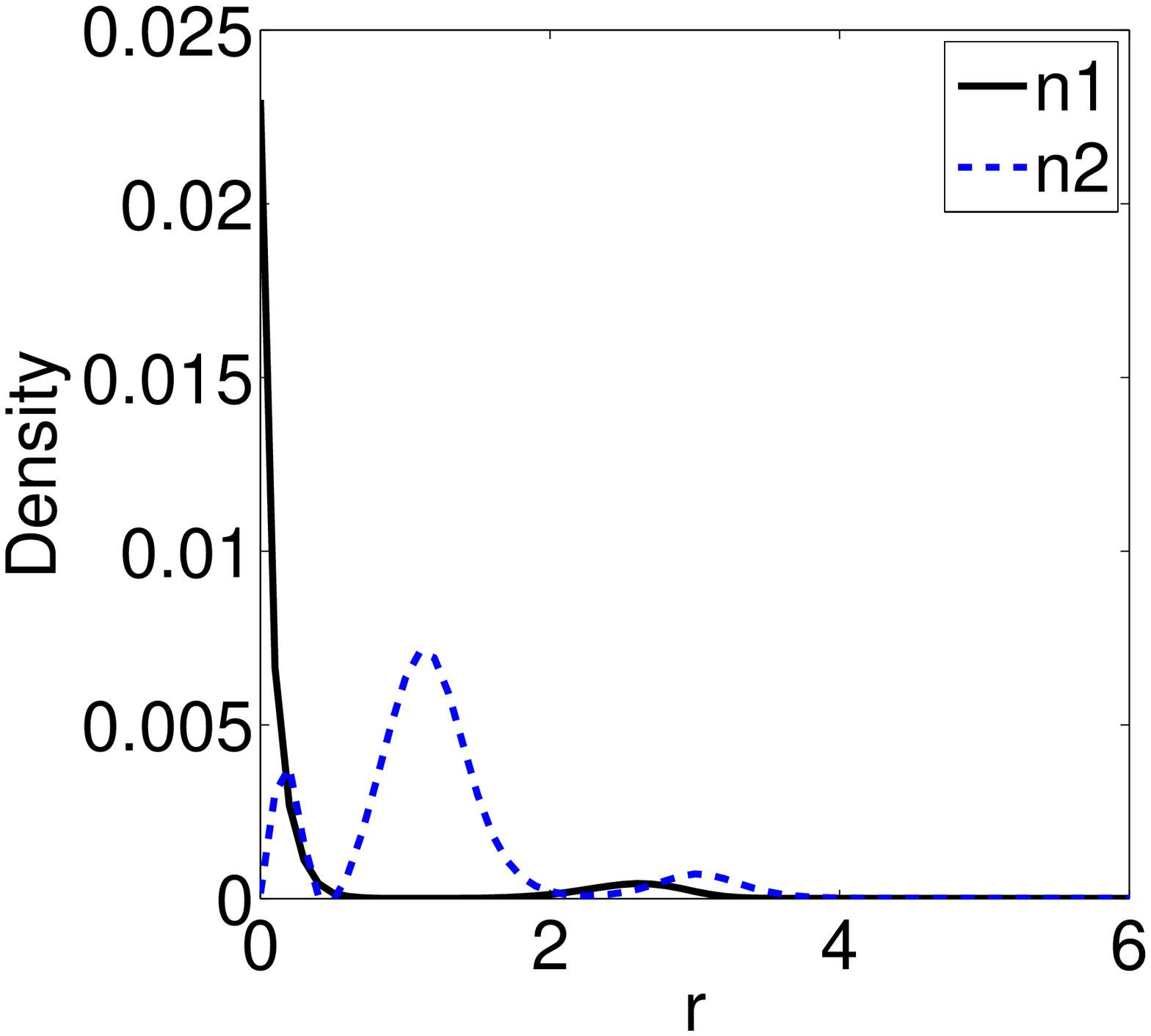}}\label{m}\hfil
\subfloat[]{\includegraphics[trim = 1cm 1.5cm 3cm 2cm,scale=.20]{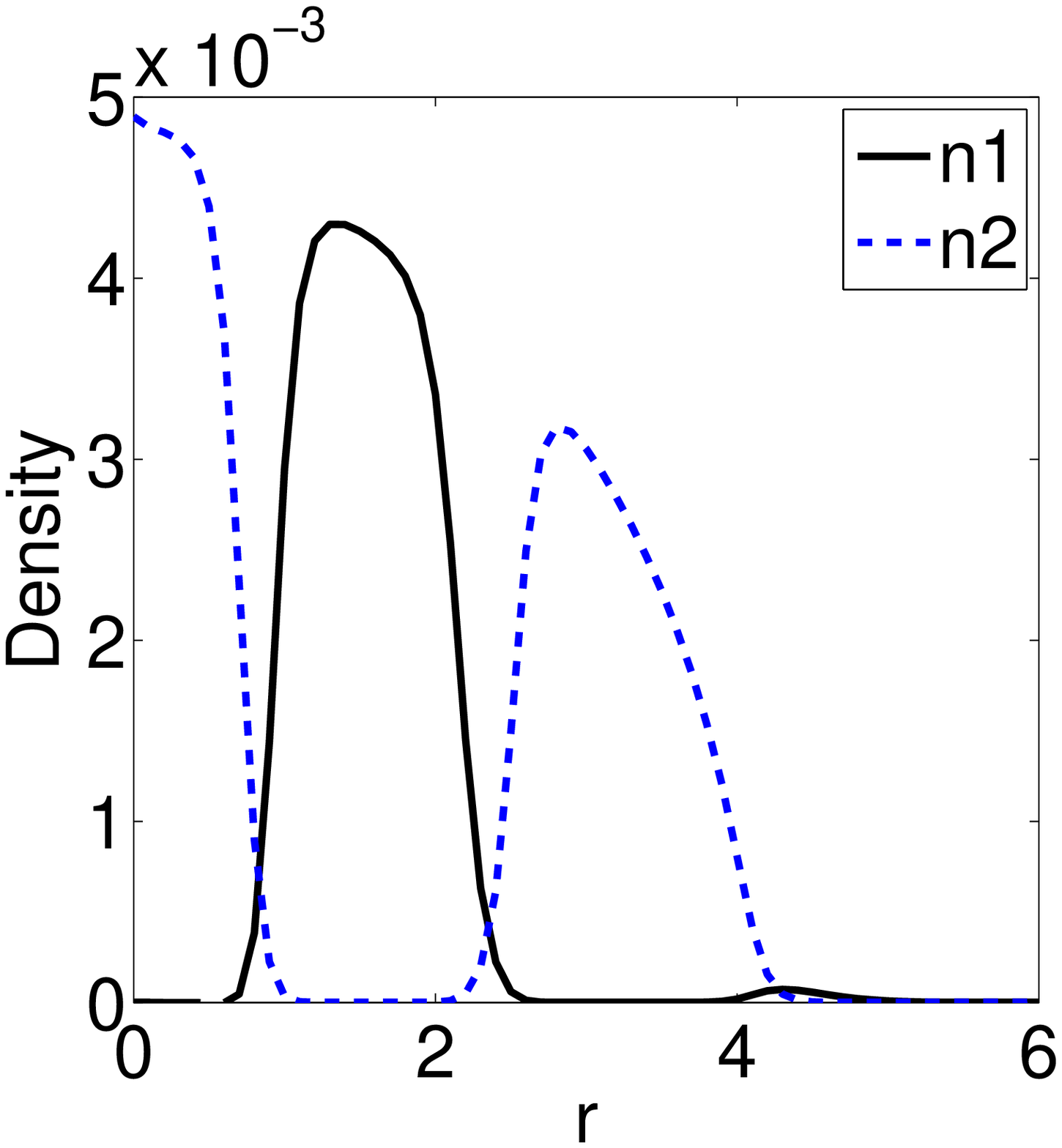}}\label{n}\hfil
\subfloat[]{\includegraphics[trim = 5cm 1.5cm 7cm 0cm,scale=.20]{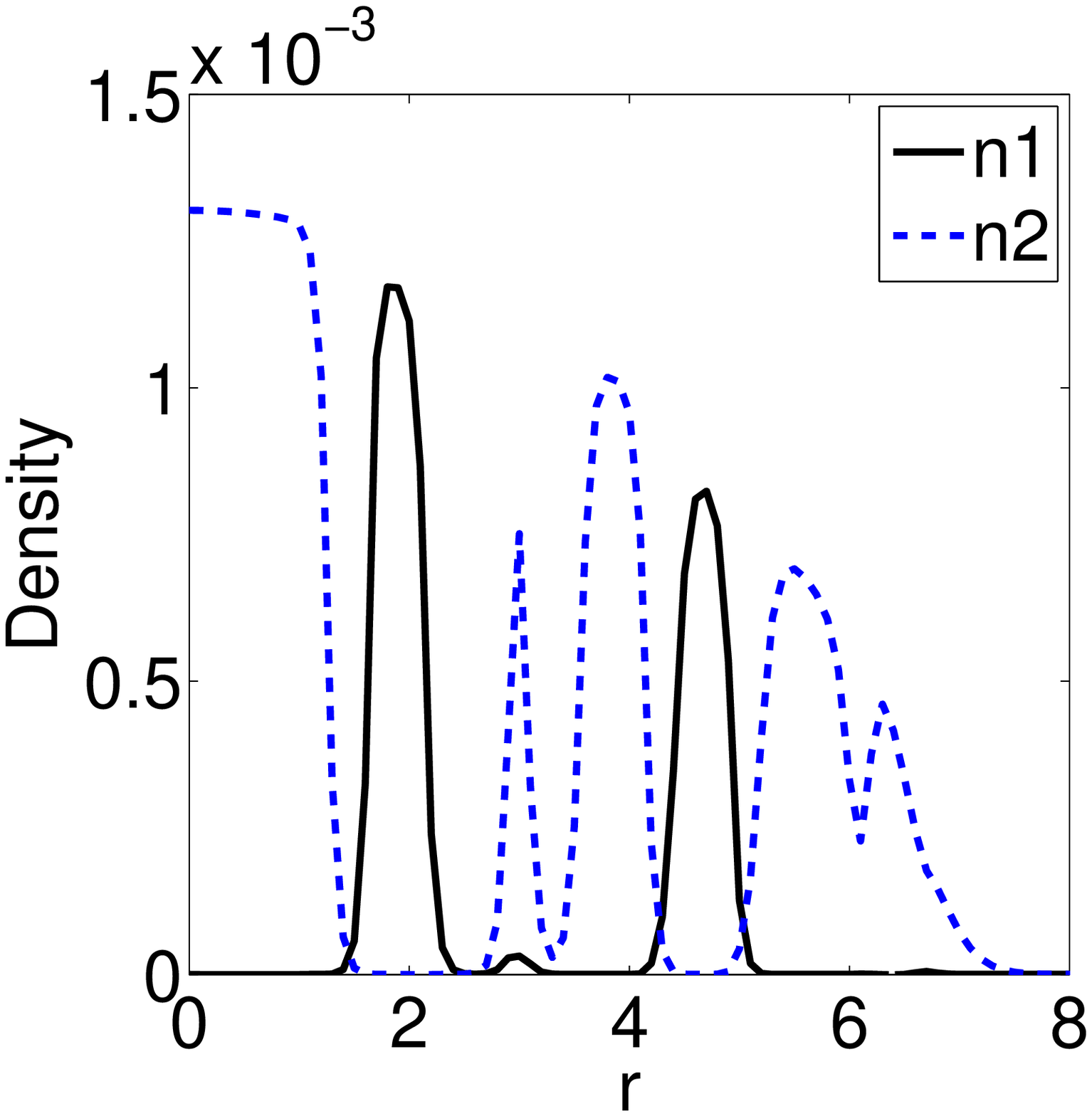}}\label{o}
\caption{Plot of the density (in unit of $a_\perp^{-3}$ ) of BEC-1 (solid lines) and BEC-2 (dotted line)  in the case of N=$10^5$ (1st column), N=$10^6$ (2nd column), and N=$10^7$ (3rd column). Rows correspond to $g=0$, $0.40$, $0.70$, $1.30$ and $1.60$. r is unit of $a_\perp$ and both the states are non-vortex states.}
\end{figure*}

\subsection{\textbf{Density profile of the components of BEC}}

Densities of BEC-1 and BEC-2 in the trap  depend on the kinetic energy of the particles, external trapping potential, intra- and inter BEC interactions.  The positive inter-BEC interaction helps  de-mixing of the BEC components, means the overlapping region will decrease with the increase in the $g$-value. Whereas, in general, positive intra-BEC mean-field interaction is proportional to the number density of the particles. This means that the span  of each of the components \cite{Dalfovo1996} increases with the increase of the density of atoms in it and this leads to more overlap in density profile among the components. Therefore, the number density of particles and inter-component interaction are anti-correlated for the overlap of the density profile of the BEC components. Further, the trapping potential tries to confine the condensates and thus also defend the phase mixing among the components \cite{Wen2013}. As a consequence, the initial density distributions of BEC-1 and BEC-2 over the extent of the trap are expected to be different for the different combination of the above parameters as discussed in the following paragraphs. As a result  of simultaneous Raman transitions on the binary components, discussed in the last section, different structures of the interference patterns will emerge.

Let's start with non-vortex BEC components with different numbers of atoms and inter-component coupling as shown in Fig. 2 corresponding to  $10^5, 10^6, 10^7$ numbers of atoms, each,  for  both the components of BEC. The plots presented here are at the level $z=0$ of the trap having azimuthal  symmetry. 
\\
\noindent
{ \bf{Case-1:}} {$g=0$: (No coupling between BEC-1 and BEC-2}) BEC-1 is more  expanded than BEC-2 as former one has relatively large scattering length.  Therefore, the central density of BEC-1 is always less compared to BEC-2 as shown  in   FIG. 2. 

\noindent
{\bf Case-2:} $0.30\leq g\le 0.90$: (mutual interaction between the BEC components is essential) Here,  many exciting features in density profiles are observed as the components start departing from each other. In all the cases, the central densities of both the components are reduced compared to  the BEC states with $g$=0 as shown in the first row of Fig. 2.    Now, when the inter-BEC interaction strength  increases, BEC-1 continues its trend of decreasing  density in the central region and eventually reaching the minimum. But BEC-2 is emerged with  opposite trend, and its density  is maximum at and near the center of the trap.  The reason behind this is  the  competition between intra- and inter-component interactions. Signature of multi-ring shaped density profile is observed for $N=10^7$.
  \\
{\bf Case-3}:  $g > 0.90$: (strong coupling between the components) As $g$ increases further,  the density of BEC-1 component is shifted away  from the center of the trap. Whereas, BEC-2 gets concentrated near the center. With increasing population in the central region, BEC-2  breaks apart, and part of it  grows at the outer surface of BEC-1 to form multi-ring shaped density profile in the x-y plane. Therefore, the fragmentation of the components is predicted for larger population as seen in  last two rows of Fig. 2.  $g$ is considered here till 1.60 and   beyond $g=1.66$  the two-component system collapses for $N=10^7$. This is the unusual situation which in general happens  for the BEC  with negative scattering length. The splitting  can also happen for the smaller number of particles, but it needs a larger inter-component interaction (not shown in figures here). Here we do see that the fragments of different components repel each other more strongly as we increase the inter-component interaction strength. This may be the reason for the collapse. Also  this critical value of $g$ changes with the trap size which will be  discussed later.

\begin{figure*}[!h]
\subfloat[]{\includegraphics[trim = 1cm 1.0cm 0.1cm 1.5cm, scale=.40]{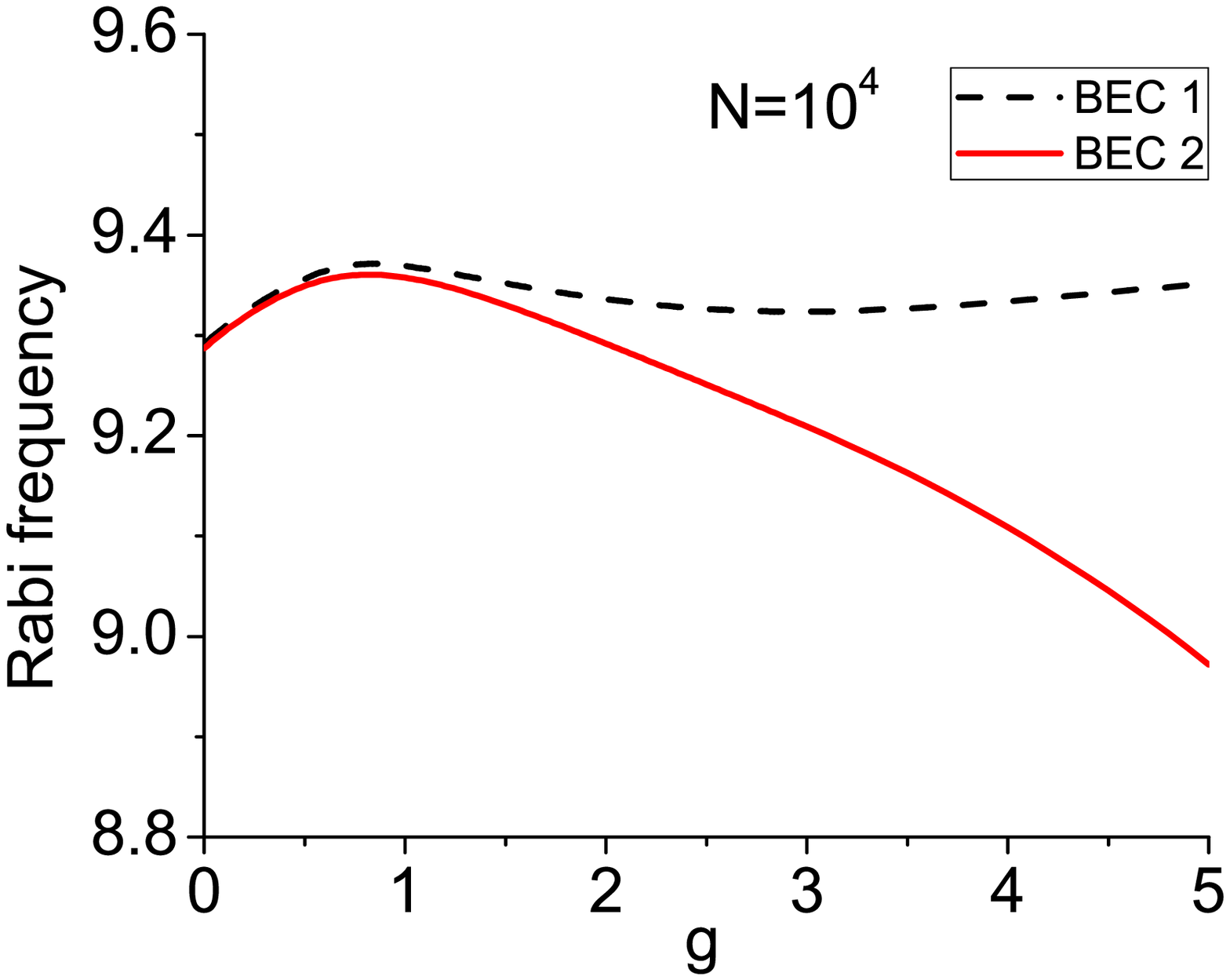}}
\subfloat[]{\includegraphics[trim = 1cm 1.0cm 0.1cm 1.5cm, scale=.40]{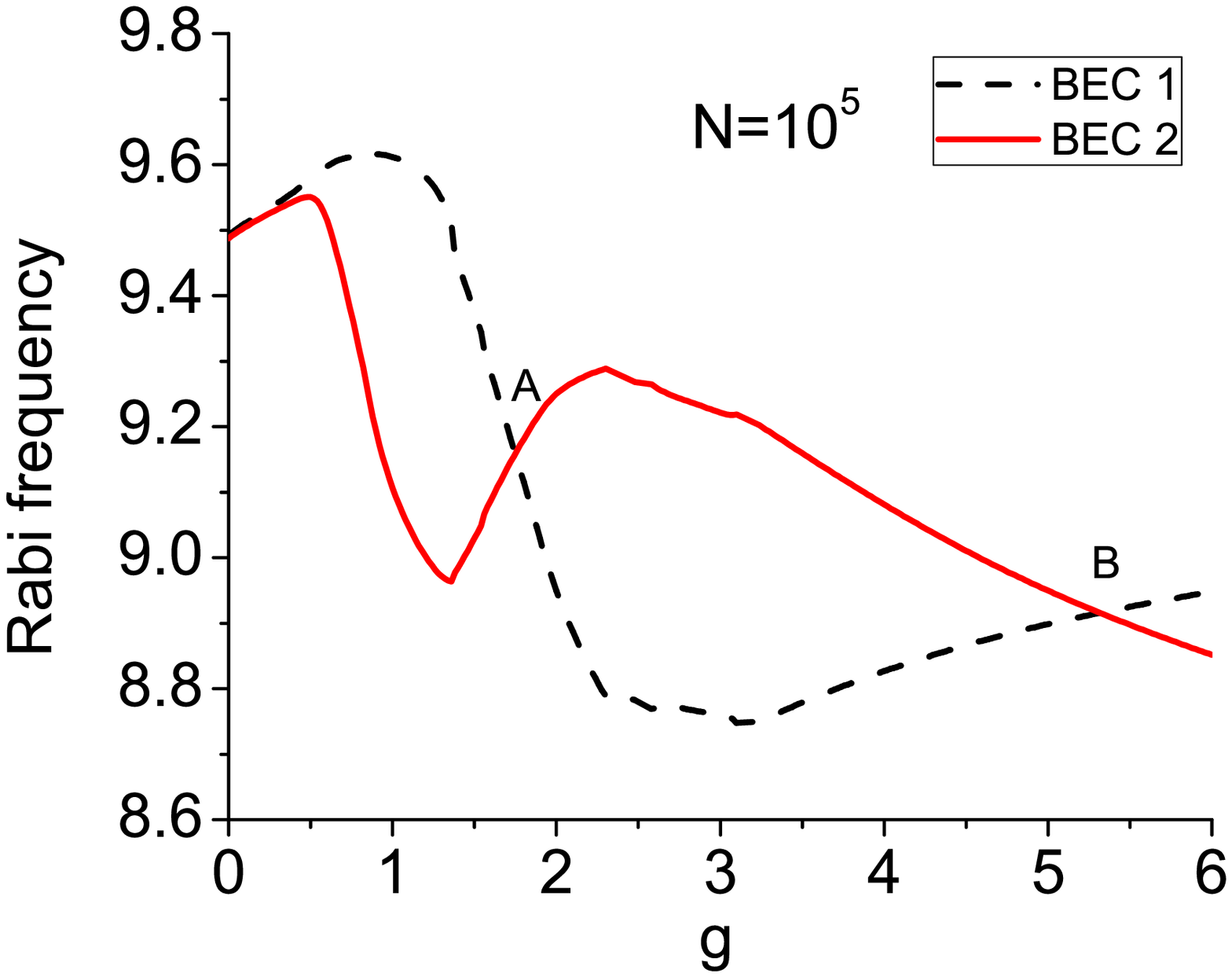}}\\
\subfloat[]{\includegraphics[trim =  1cm 1.0cm 0.1cm 0.1cm,scale=.40]{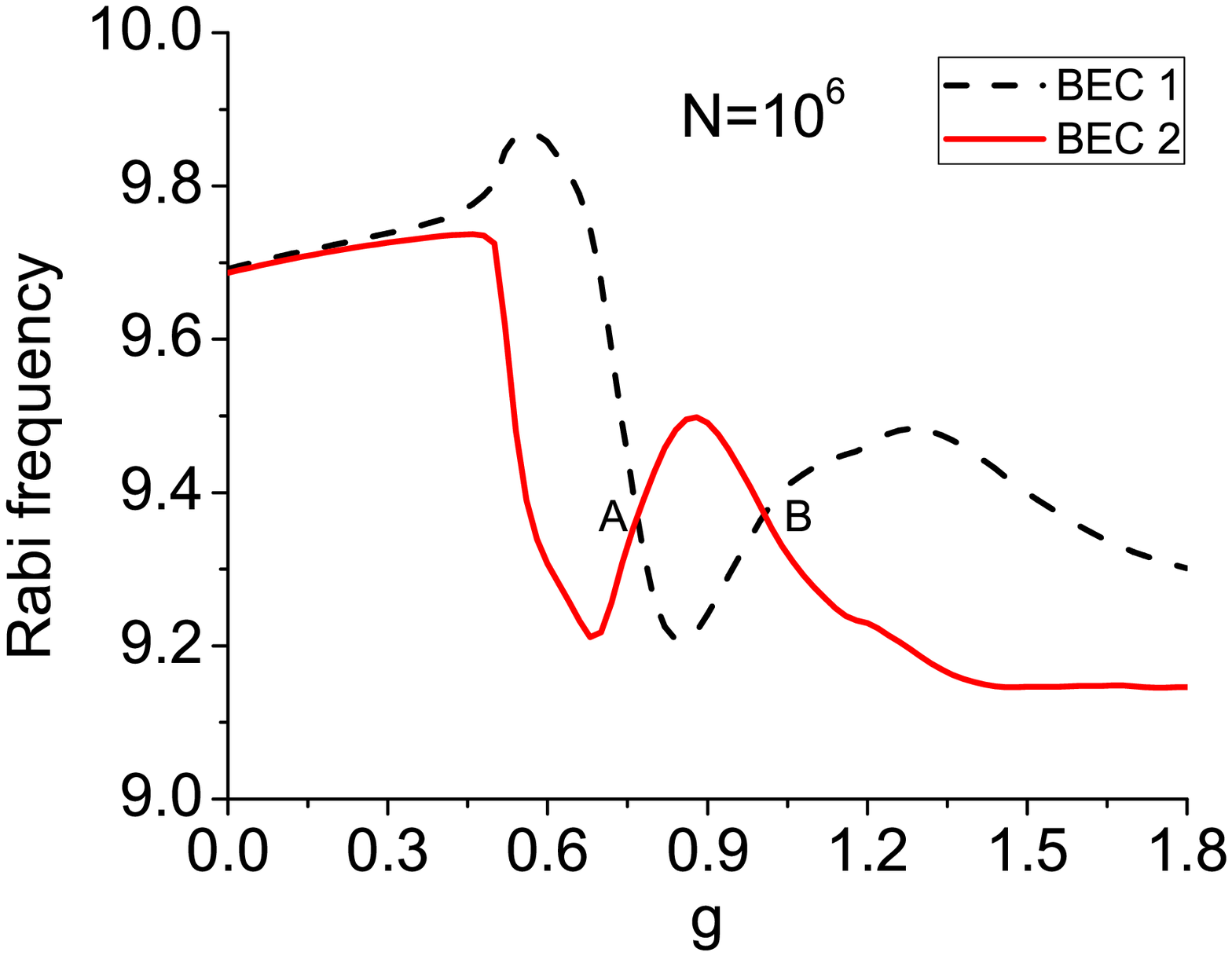}}
\subfloat[]{\includegraphics[trim =  1cm 1.0cm 0.1cm 0.1cm, scale=.40]{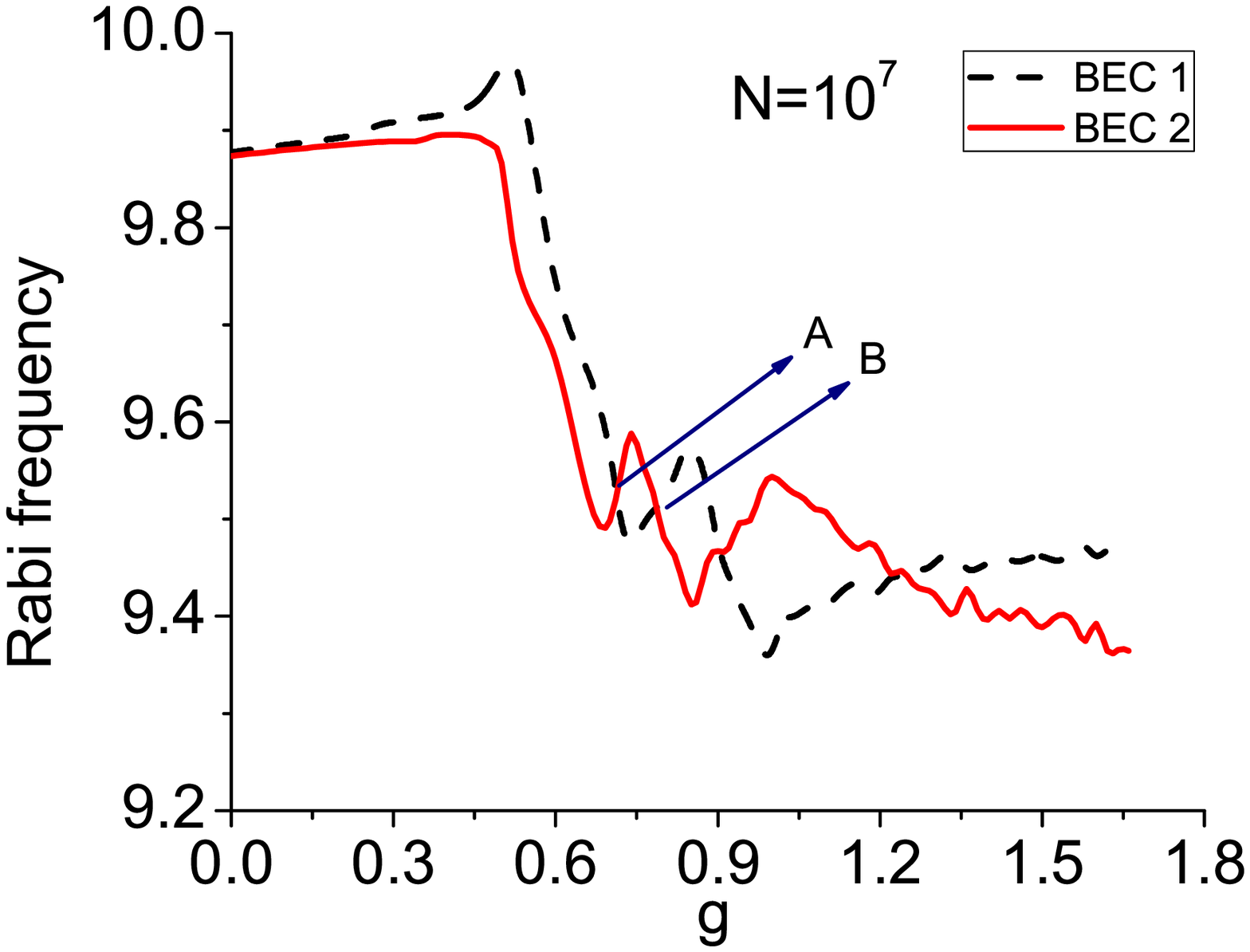}}
\caption{Variation of dipole Rabi frequency (in Sec$^{-1}$) of BEC-1  and BEC-2 with $g$ on a semilog scale. LG beam of OAM=+1 interacts with non-vortex BECs.}
\end{figure*}

\begin{figure*}[!h]
\subfloat[]{\includegraphics[trim = 1cm 1.0cm 0.1cm 0.1cm, scale=.40]{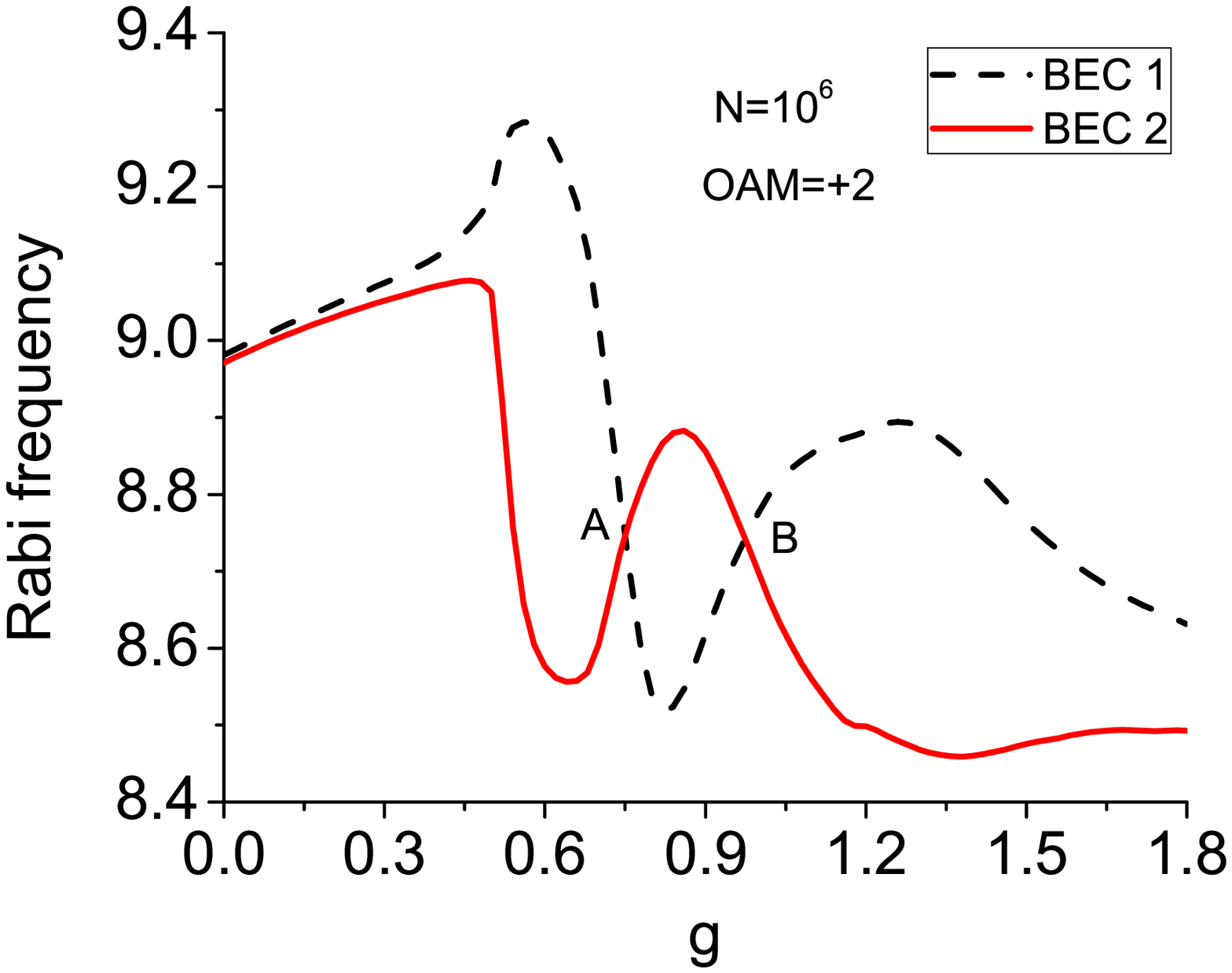}}
\subfloat[]{\includegraphics[trim = 1cm 1.0cm 0.1cm 2cm, scale=.40]{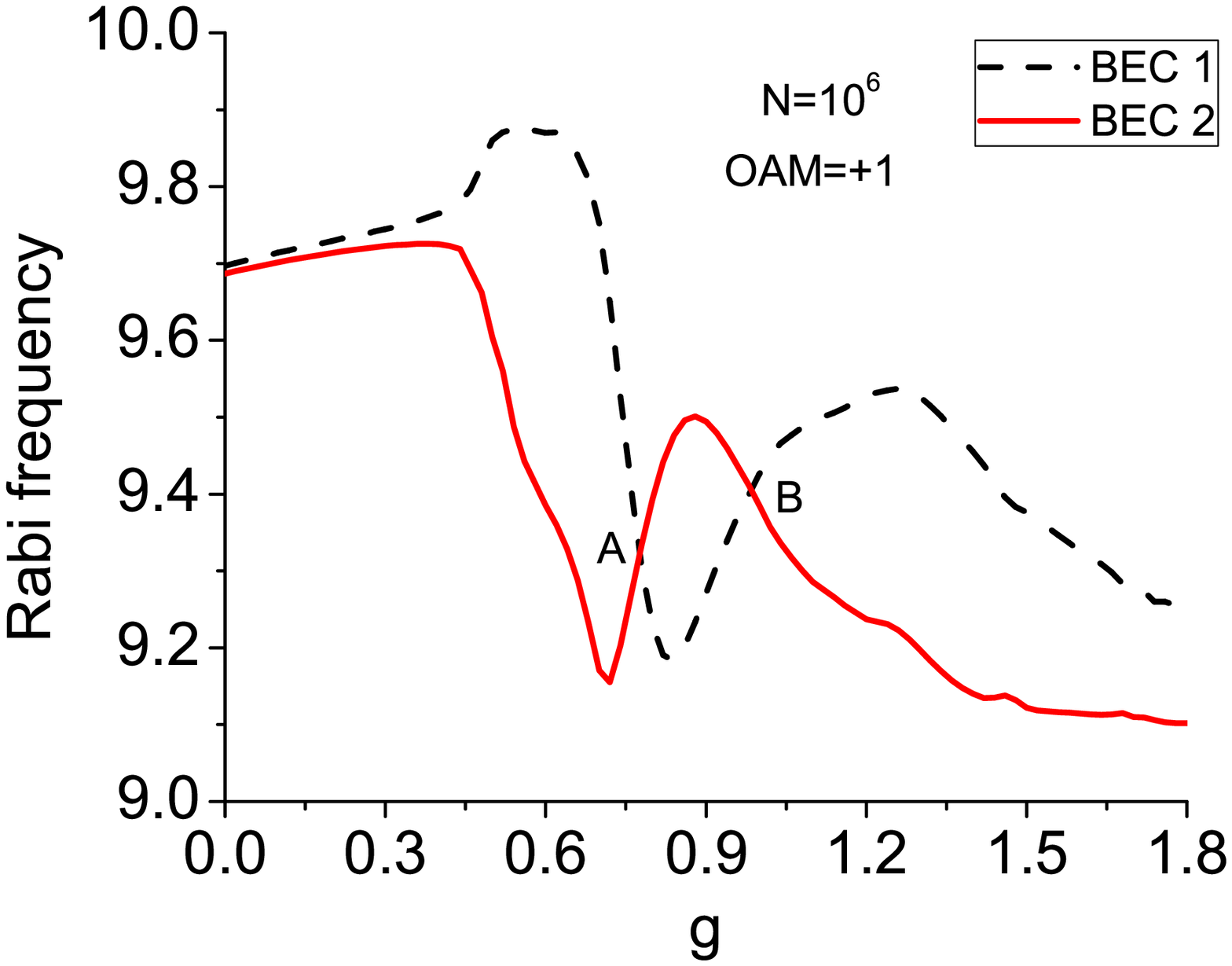}}\\
\subfloat[]{\includegraphics[trim = 1cm 1.0cm 0.1cm 0.1cm, scale=.40]{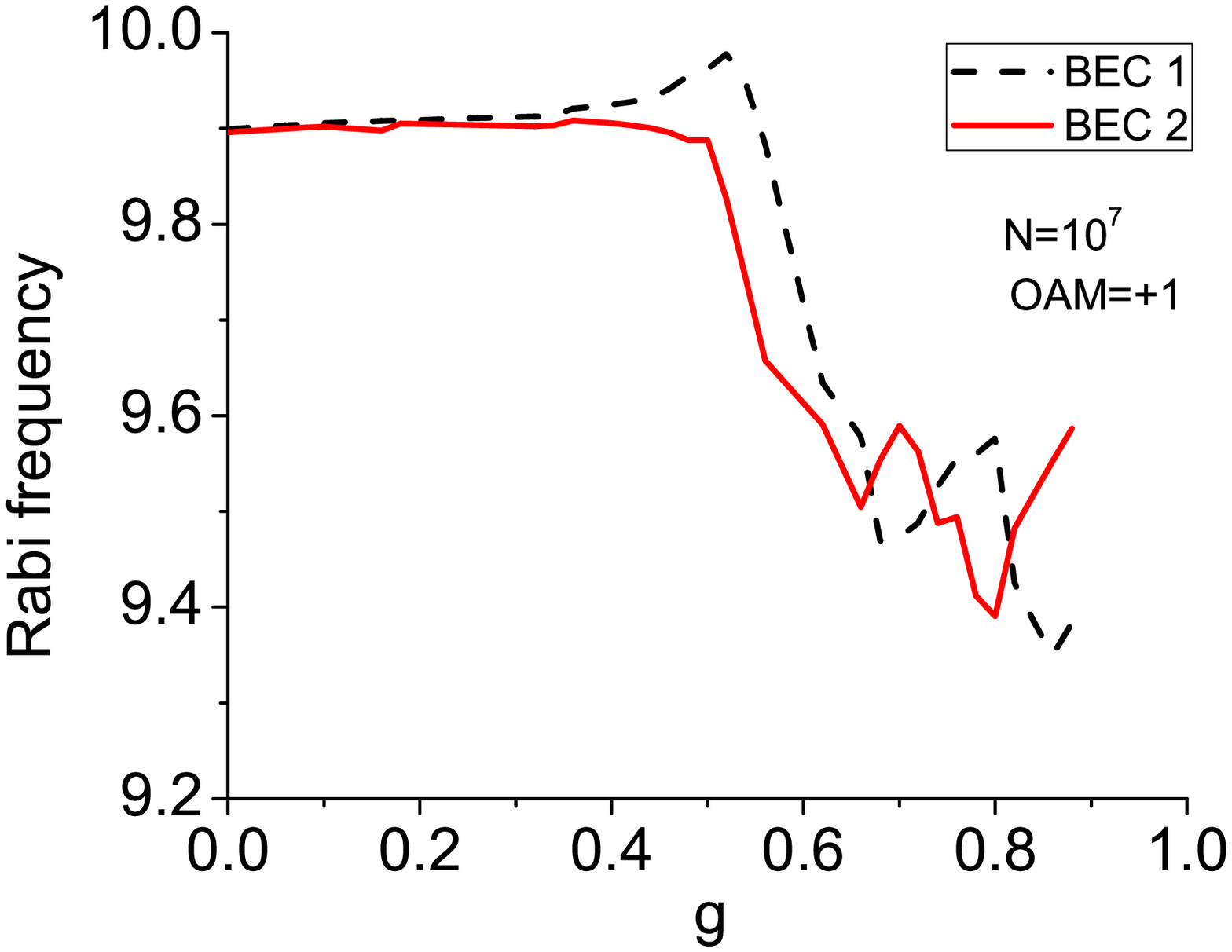}}
\subfloat[]{\includegraphics[trim = 1cm 1.0cm 0.1cm 2cm, scale=.40]{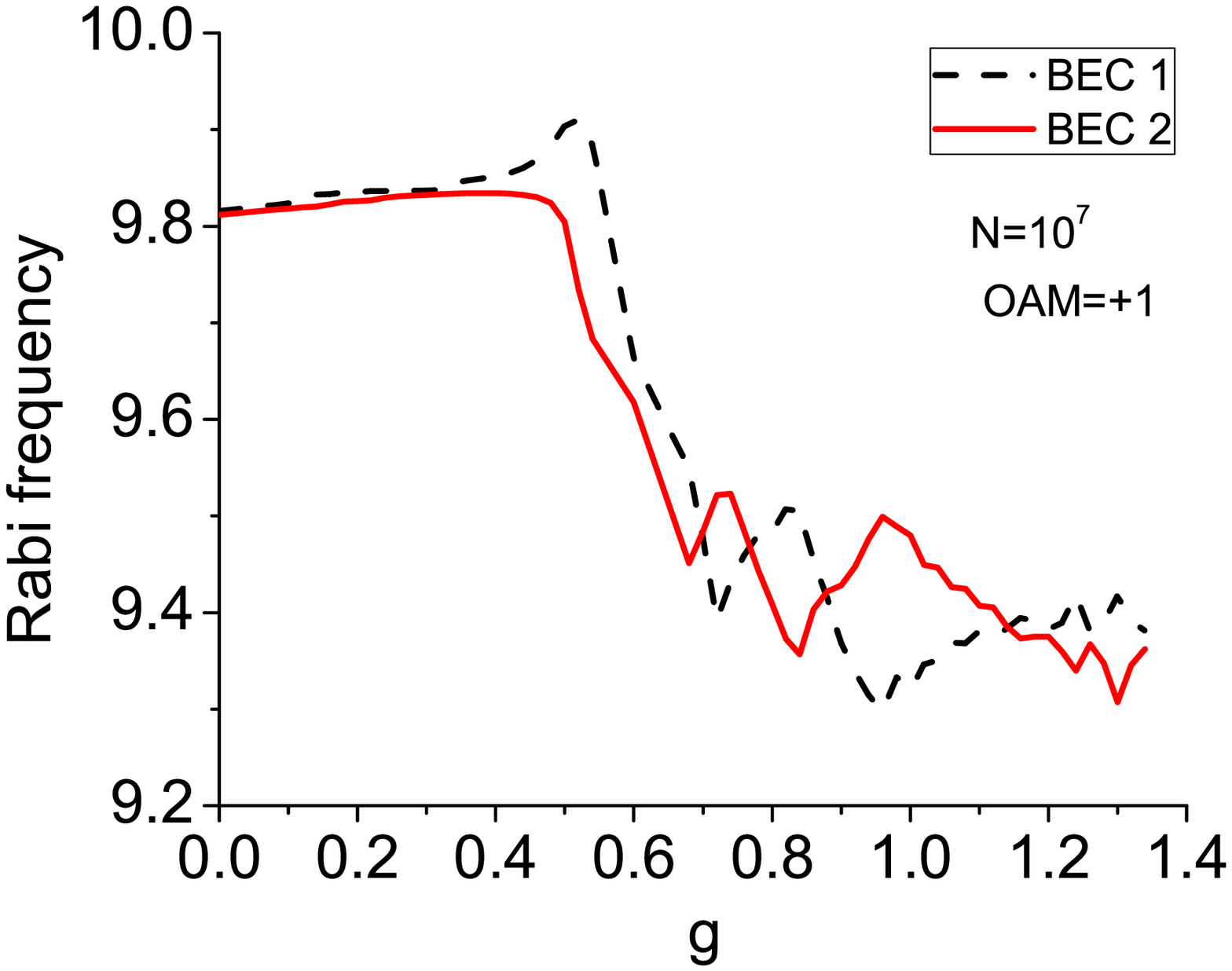}}

\caption{Variation of dipole Rabi frequency (in Sec$^{-1}$) of BEC-1  and BEC-2 with $g$ on a semi-log scale. LG beam  interacts with non-vortex BECs with $N=10^6$ for (a) $a_{11}=1.03\times 5.5$nm, $a_{22}=0.97\times 5.5$nm and (b) $a_{11}=1.06\times 5.5$nm, $a_{22}=0.94\times 5.5$nm.  For $N=10^7$ with $a_{11}=1.03\times 5.5$nm, $a_{22}=0.97\times 5.5$nm and (c) $a _\bot =3$  $\mu$m and (d)$a _\bot =4$  $\mu$m. }

\end{figure*}

\subsection{\textbf{Interaction of BECs with LG beam}}

Now we would like to discuss the interaction of these two-component BECs with the LG beam having topological charge +1, say (physics will be similar for -1 also). FIG. 3 presents the variation in Rabi frequencies  of two-photon Raman transitions as described in FIG. 1,  with respect to  the $g$-value.  All the density profiles  used in the last section have been used here for initial wavefunctions of the components of BEC. Variations of the Rabi frequencies have been presented in the figure to show consistent behavior with other large populations at low $g$-values and to confirm no unusual feature for moderate or large $g$-values. The plot confirms that the BEC-2 component shrinks in the trap with the increasing inter-component interactions. For populations $10^5$  or more,  density profiles of BEC start showing unusual behaviour starting from $g=0.5$.  The components of BEC peak at different places, something like peaks of the BEC-1 and BEC-2 appear alternately.  Therefore, the interaction of the BEC with the LG beam for these ranges of parameters is expected to provide interesting physics in terms of Rabi frequencies.

Having larger intra-component interaction, BEC-1 has relatively larger Rabi frequency than BEC-2 in the region of weak  inter-component coupling  (i.e.$g\leq 0.5$). This is  due to the larger overlap of the beam profile with BEC-1 than BEC-2. Both the Rabi frequencies are increasing initially  in this region of $g$ because  of the uniform enlargement of the span of matter density and it accelerates with the population of atoms. 

The initial reduction of  Rabi frequencies of BEC-2 is consistent with the fast collapse of its density over radial span  with increasing $g$-values. Further, the density collapse is estimated faster with respect to $g$-values as total population in the system increases. The trend is similar for BEC-1, though its collapse occurs at relatively larger values of $g$ as its density profile has stronger overlap with beam profile up to the relatively larger values of $g$. Interestingly, the separation between the collapse regions of $g$-values for both the components is narrowing down with the increase in total populations and almost overlap around 0.5 to 0.7 for $N=10^7$. 

This can be understood if we compare the density profiles of the components around the radial region of  two times the characteristic length of the trap, where the intensity profile of the LG beam is also  significant.

  One of the striking features of the distribution of Rabi frequencies displayed in FIG. 3, is that  the hierarchy of strength of Rabi frequencies between BEC-1 and BEC-2 is exchanged at certain values of  $g$. These degenerate  points in the plots  are denoted  as 'A' and 'B'. The separation of A and B points is reduced with the increase of the number of particles.  In FIG 3 (d), the hierarchy between the Rabi frequencies  gets phase changed multiple times within  $g=2$. Here the arbitrariness  of the lengths  between 'A' and 'B'-type critical points  is because of the instabilities of the BECs at higher $g$-values.

  To understand the dependency of the distribution on different parameters of the beam, trap geometry and properties of atoms, we  consider the cases where a particular parameter is  changed keeping all the other parameters fixed as discussed for FIG. 3. FIG. 4 displays the variation of Rabi frequencies over the range of $g$.  FIG. 4(a) shows variation of Rabi frequencies for  $N=10^6$, OAM=+2. It is clear that the crossing points, A and B,  of the distribution, have a negligible dependence on the charge of the optical vortex when it is compared with FIG 3(c). But the maximum difference between the distributions in between the points A and B has increased in the latter case, and this can reduce the visibility of the interference pattern discussed in the next section.  FIG. 4(b) displays the same variation as FIG 3(c), but  $(a_{11},a_{22})=(1.06,0.94)\times 5.5nm$. Here we see significant effect compared to FIG 3(c) where $(a_{11},a_{22})=(1.03,0.97)\times 5.5nm$ is used.  
  FIG. 4(c) and FIG. 4(d) show the  fluctuations of Rabi frequencies for tighter  trapping potentials with
$a_\bot=3 \mu m$ and $a_\bot = 4 \mu m$, respectively, in case of $N=10^7$. We observe multiple crossings  between the spectrum of Rabi frequencies for BEC-1 and BEC-2. As expected, the components of BEC are collapsed at lower $g$-values: 0.88 and 1.34, respectively.

\section{CREATION OF VORTEX-ANTIVORTEX STATES WITH EQUAL AND UNEQUAL QUANTUM CIRCULATION}

Formation  of the quantized vortex and antivortex, and their superposition  in the BEC by the LG beam have been experimentally studied \cite{Andersen2006,Wright2008}  over last decade to understand the properties  of vortices in BEC. The coherent superpositions of  vortex-antivortex of equal or unequal circulation quantum numbers  \cite{Wright2008, Wright2009} yield interesting interference effects with potential applications \cite{Brachmann2011, Quinteiro2014}, such as manipulating the chirality  \cite{Mondal2015, Bhowmik2016, Toyoda2012}. In these studies, the matter-wave vortex is shown to  acquire vorticity  equal to the winding number of the LG beam corresponding to  electronic dipole transition.

To make this superposition, let us consider that the LG beam  with positive  vorticity interacts with BEC-1 and negative vorticity with BEC-2 as shown in FIG.1. The two-photon Raman transitions produce  vortex-antivortex pair in the hyperfine state $| \psi _i \rangle=| 5S_{\frac{1}{2}}, F=2, m_f =0 \rangle$. The interference pattern of the superposition will depend on the populations of the vortex states. Thus  the Rabi frequencies corresponding to these  two-photon transitions are important for the coherency of the interference pattern.

In general, the two different macroscopic vortices  with vorticities $l_1$, $l_2$ superpose  with arbitrary proportion  as \cite{Bhowmik2016,Liu2006}

\begin{equation}
\Psi(R,\Phi, Z, t)= f(R,Z) e^ {-i\mu t}(\alpha_1 e^{il_1\Phi} + \alpha_2 e^{il_2\Phi} ),
\end{equation}
 where $R^2=(X^2+Y^2)$, $\mu$ is chemical potential of the system.  The constants, $\alpha_1$ and $\alpha_2 $, depend on the strengths of two-photon transitions corresponding to vortex and antivortex, respectively,  with $ |\alpha_1 |^2 +| \alpha_2 |^2 =1$. All the density structures presented in Fig. 5 are at $Z=0$ plane.  The left and right columns of FIG. 5 present the density profile of  $N=10^6$ BEC at vortex-antivortex superposed state for $g=0.76$ and $g=0.86$, respectively . Here, choice of quantum circulation  are ($l_1,l_2$) = (1, -1) for 5(a) and 5(b),  ($l_1,l_2$) = (1, -2) 
 or (2,-1) for 5(c) and 5(d), and ($l_1,l_2$) = (2, -2) for 5(e) and 5(f). The choice of  $g= 0.76$ is considered here for $|l_1|=|l_2|$ to show the  coherent interference where the populations of  both the components are equal. We  get same interference pattern also at the point 'B'  as well as at $g=0$ where there is no interaction  among the components. In the same spirit,  $g=0.86$ (between A and B points) is chosen where maximum deviation of Rabi frequencies occurs.   First and third rows of FIG. 5 show symmetric fringe patterns with respect to X and Y axes due to equal amplitudes of $l_1$ and $l_2$.  Whereas, second row is one of the examples of superposition of vortex and antivortex with unequal quantum circulation number. Also, the observation of distributions of Rabi frequencies for BEC-1 and BEC-2 ( FIG. 3(c) and  FIG. 4(a)) justify why the visibility in this case is much better for $g=0.86$ than $g=0.76$ for  unequal quantum circulation number of vortex-antivortex.  In all these interpretations, we keep in mind  that   approximately equal number of particles with opposite orientation can produce clear fringe pattern.   In case of non-equal initial populations among BEC-1 and BEC-2, we will get similar patterns of interference at different values of $g$.
 
 \begin{figure*}[!h]
\subfloat[]{\includegraphics[trim = 2cm 1cm 6cm 2cm, scale=.24]{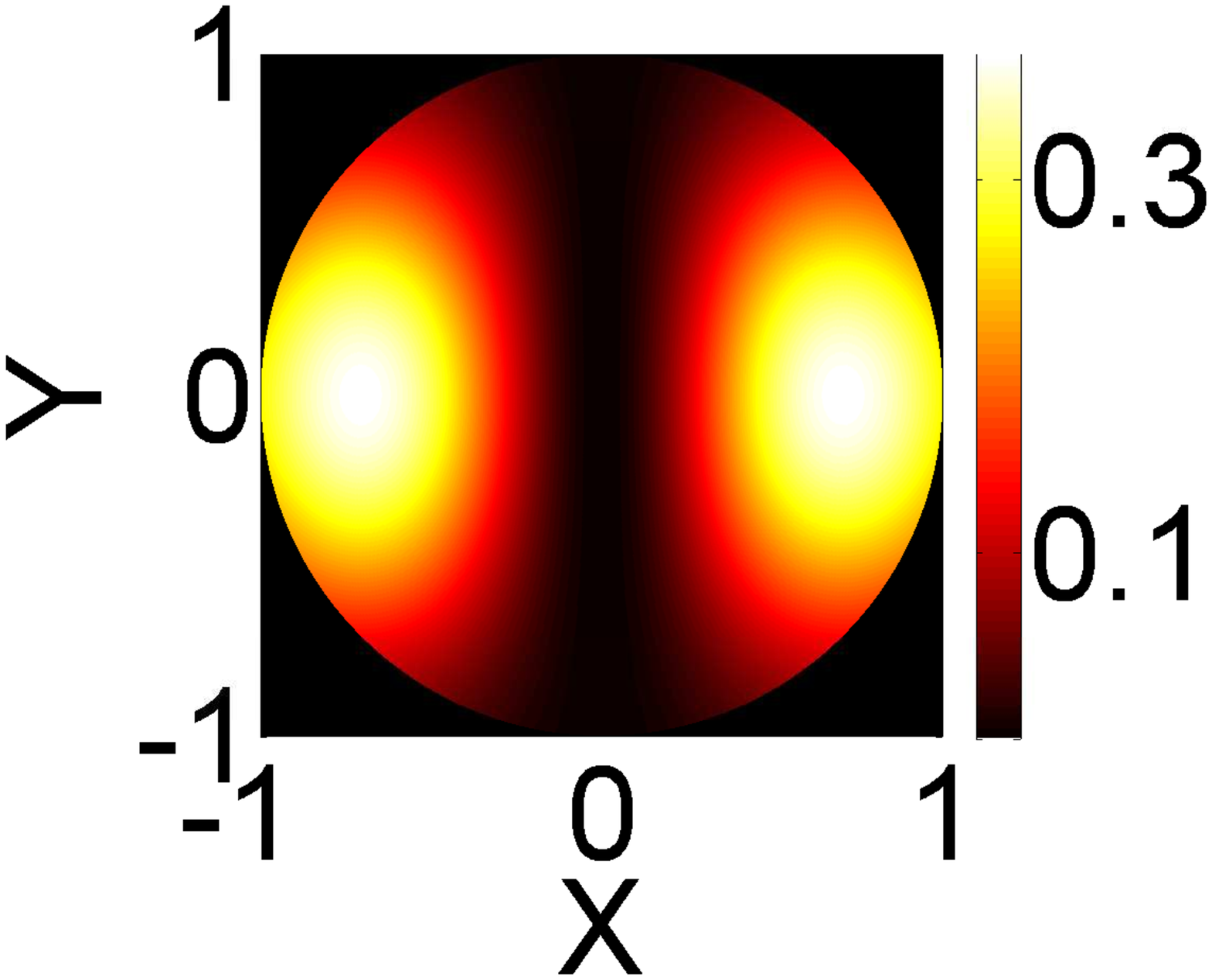}}
\subfloat[]{\includegraphics[trim = 1cm 1cm 5cm 2cm, scale=.24]{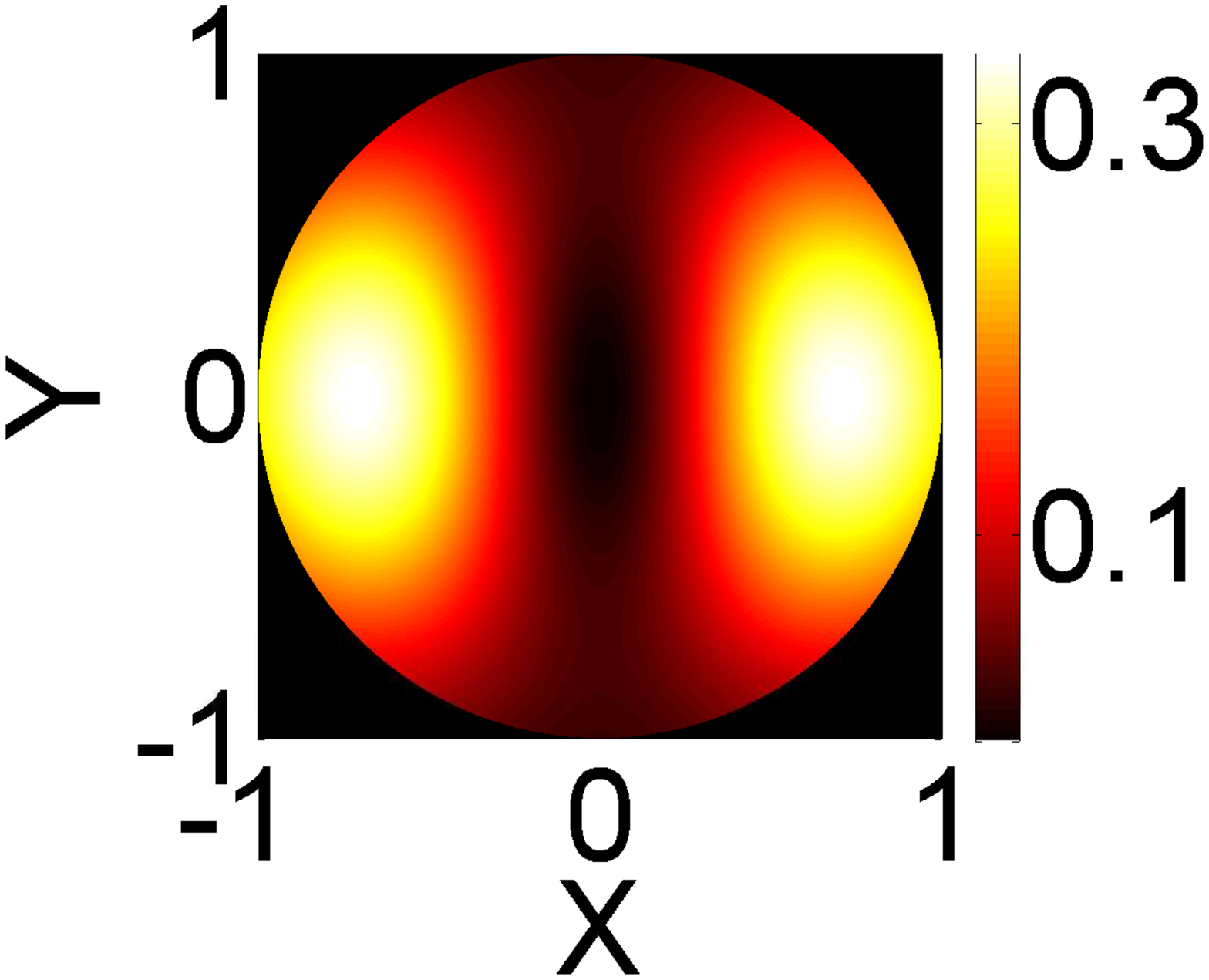}}\\
\subfloat[]{\includegraphics[trim = 2cm 1cm 7cm 2cm,scale=.24]{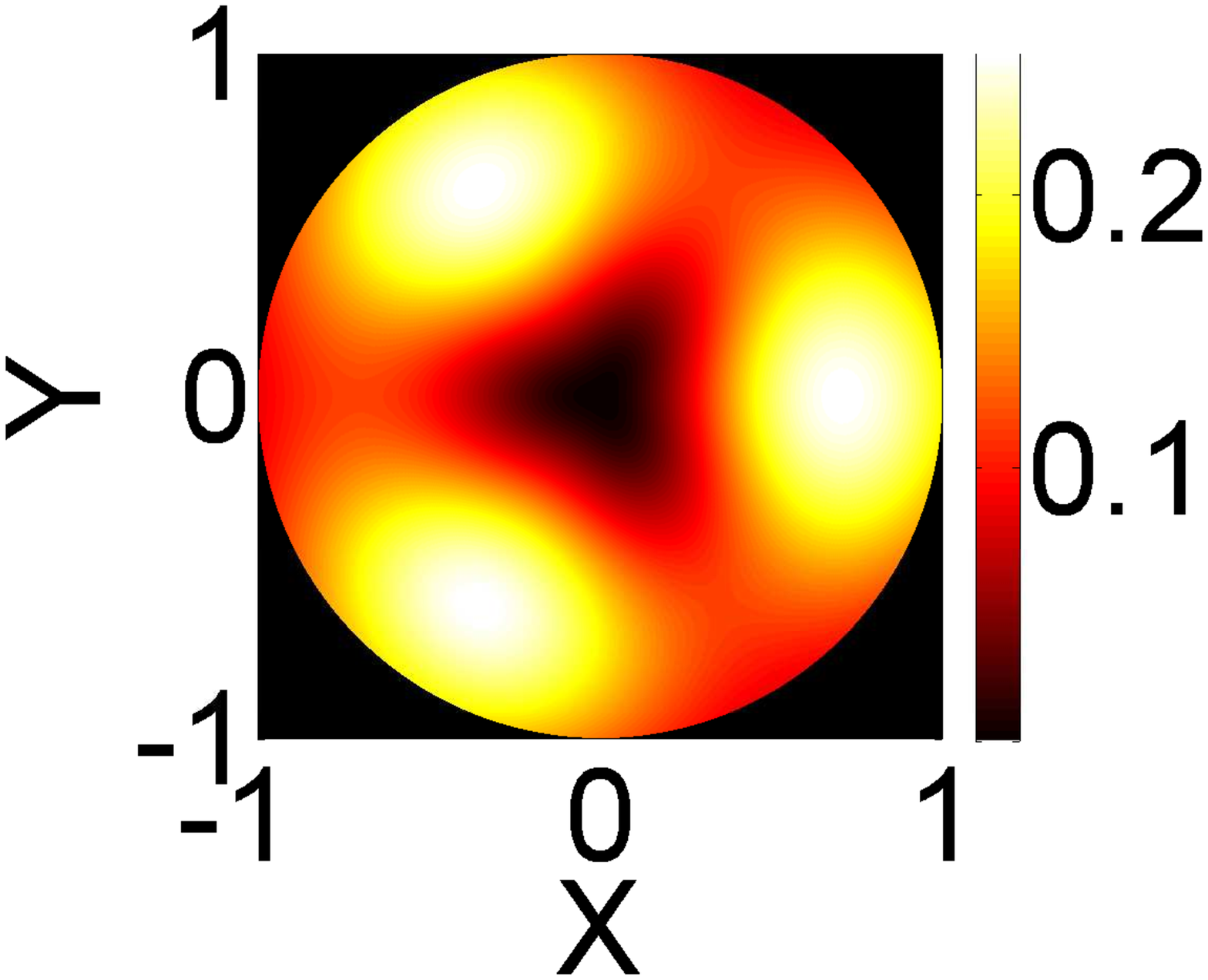}}
\subfloat[]{\includegraphics[trim = 1cm 1cm 5cm 2cm, scale=.24]{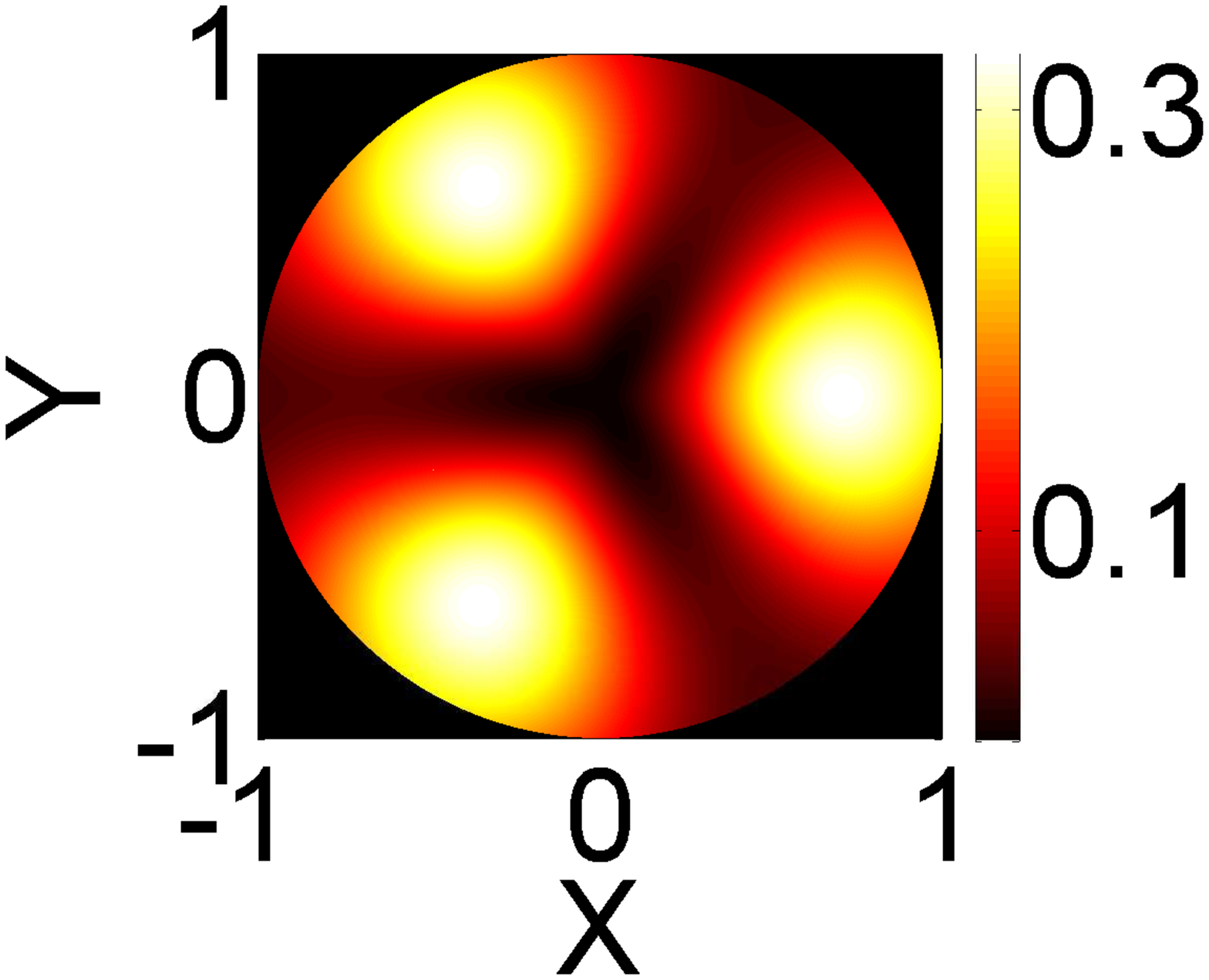}}\\
\subfloat[]{\includegraphics[trim = 2cm 1cm 7cm 2cm, scale=.24]{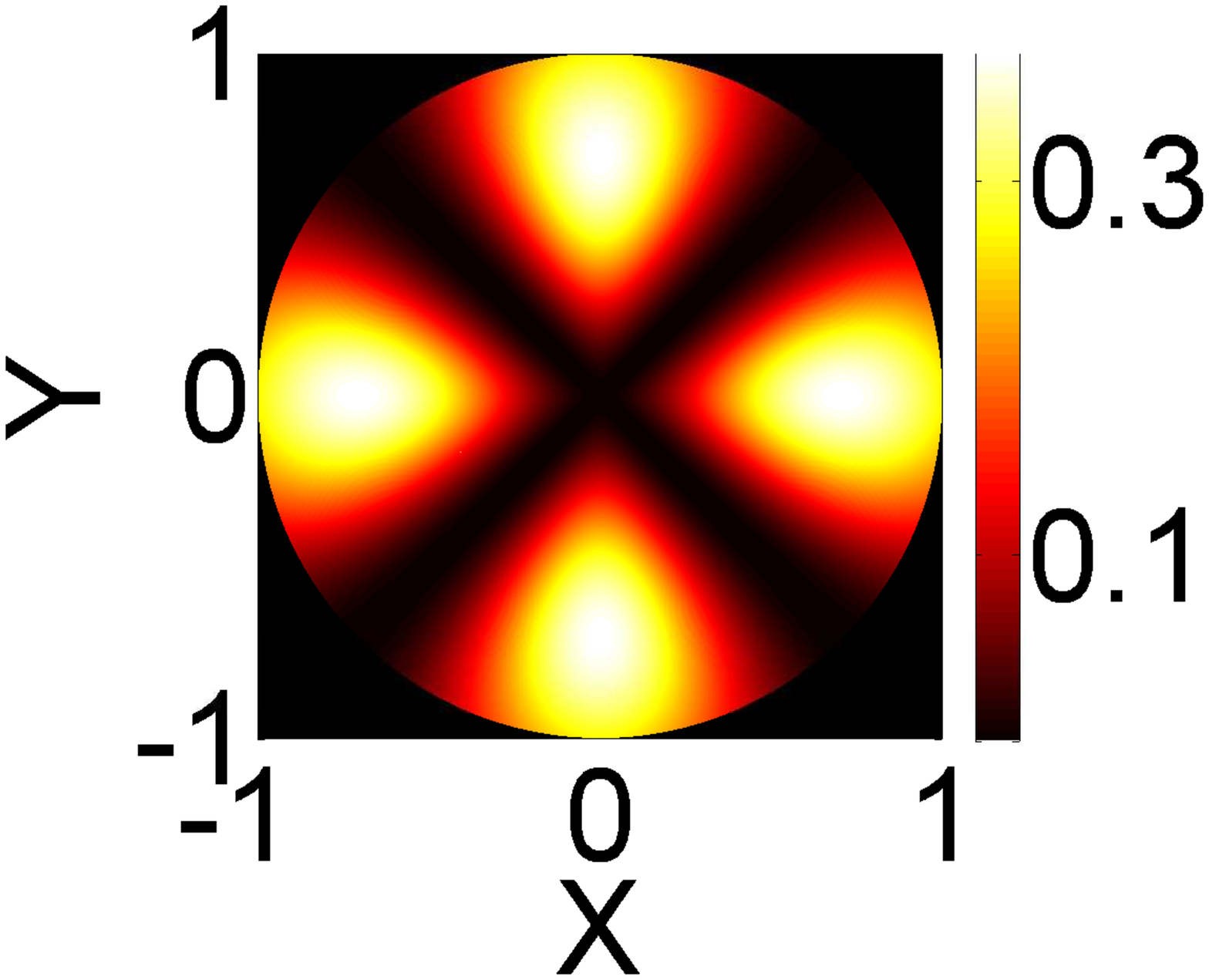}}
\subfloat[]{\includegraphics[trim = 1cm 1cm 5cm 2cm,scale=.24]{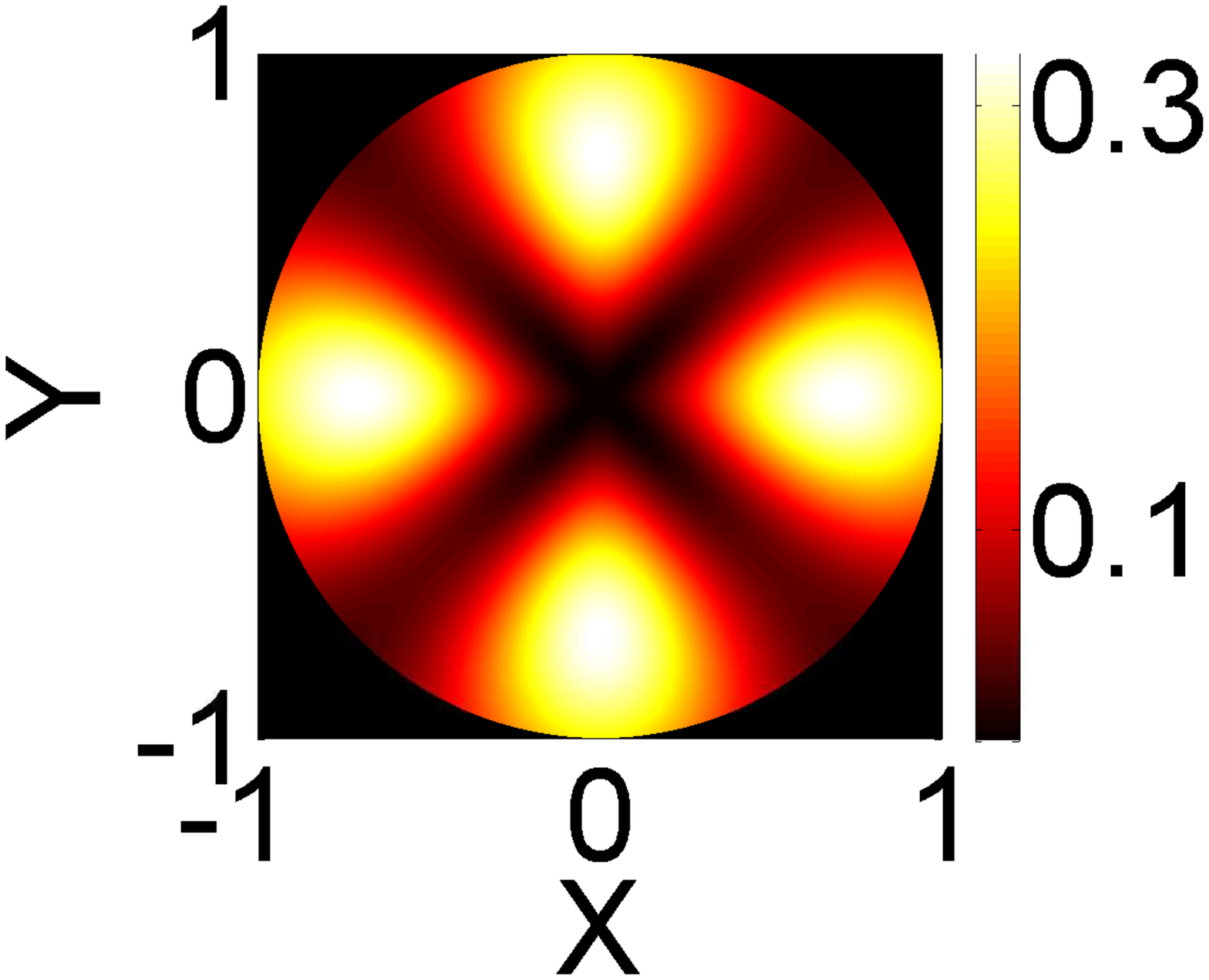}}\\

\caption{Plot of  the density distribution of  vortex-antivortex states of $N=10^6$ with ($l_1,l_2$) = (a)\& (b)(1, -1), (c) \& (d) (1, -2), (e) \& (f) (2, -2). Left and  right columns are for $g=0.76$ and  $g=0.86$ respectively. All quantities are in dimensionless units.}
\end{figure*}

\section{CONCLUSION}

We have developed a theory of interaction of LG beam with binary mixtures of  BEC and have shown the variability  of the vortex-antivortex superposed state. Competition between intra- and inter-BEC  interactions for the two-component non-vortex ground state has been shown in the graphical representation. The effects of the number of particles in this binary condensation  have also been investigated.    For $N=10^7$, the critical value of $g$ has been found out, for which the BECs  collapse. The effects of the trapping potential on the critical values of $g$  have also been studied along with the degeneracy points of Rabi frequencies. We have shown,  how the wavefunctions of the initial states of the components directly affect the  Rabi frequency  of the two-photon stimulated Raman transition. Here  the two photons of Raman transition are made of LG and G beam. The calculated Rabi frequencies help us to find out  the  population density of the vortices of each of the components. These changes in population density of  vortex-antivortex states  show the variation of  interference patterns.

\section*{ACKNOWLEDGMENTS}
The calculations were performed in the IBM cluster at IIT-Kharagpur, India funded by DST-FIST.

\section*{APPENDIX}
To derive the coupled-GP equations, we start from the action functional  \cite{Kevrekidis2008}
\begin{equation}
S=\int(\mathcal{L}_1+\mathcal{L}_2-U_{12}|\Psi_1|^2 |\Psi_2|^2)d^3\textbf{r} dt,
\end{equation}

where the Lagrangian density of each component is
\begin{equation}
\hspace{-2.3cm}\mathcal{L}_k=i\frac{\hbar}{2}\left(\Psi_k^*\frac{\partial}{\partial t}\Psi_k-\Psi_k\frac{\partial}{\partial t}\Psi_k^*\right)-\frac{\hbar^2}{2m_k}|\nabla \Psi_k|^2-V_k|\Psi_k|^2-\frac{\kappa^2}{R^2}|\Psi_k|^2-\frac{U_{kk}}{2}|\Psi_k|^4.
\end{equation}
Here $ U_{kk} $ is the intra-atomic coupling strength and $V_k$ is the external trapping potential.

Now, the coupled GP equations can be obtained by extremizing the action $S$ with respect to $\Psi_k^*$ using the equation  $ \frac{\partial S}{\partial \Psi_k^*}=0 $ \cite{Kevrekidis2008}. Since the trapping potential $V_k=\frac{1}{2}m_k(\omega^2_{\bot}R^2+\omega^2_{Z}Z^2)$, then the two-component GP equations take the form as 

\begin{equation}
\hspace{-2.3cm}\left[-\frac{\hbar^2\nabla^2}{2m_1}+\frac{1}{2}m_1(\omega^2_{\bot}R^2+\omega^2_{Z}Z^2)+\frac{\kappa^2}{R^2} + U_{11}|\Psi_1|^2+ U_{12}|\Psi_2|^2\right]\Psi_1=\mu_1 \Psi_1,
\end{equation}

\begin{equation}
\hspace{-2.3cm}\left[-\frac{\hbar^2\nabla^2}{2m_2}+\frac{1}{2}m_2(\omega^2_{\bot}R^2+\omega^2_{Z}Z^2)+\frac{\kappa^2}{R^2}+ U_{22}|\Psi_2|^2+U_{21}|\Psi_1|^2\right]\Psi_2=\mu_2 \Psi_2.
\end{equation}

\clearpage

% ============================================================================
% === REFERENCES =============================================================
% ============================================================================

%\bibliographystyle{abbrv}

%\bibliography{bib}

\end{document}